\documentclass[review,12pt]{elsarticle}

\usepackage{array}
\newcolumntype{L}[1]{>{\raggedright\arraybackslash}p{#1}}
\newcolumntype{C}[1]{>{\centering\arraybackslash}p{#1}}

\PassOptionsToPackage{expansion=false}{microtype} 
\PassOptionsToPackage{hidelinks,breaklinks,unicode}{hyperref} 

\usepackage[utf8]{inputenc}
\usepackage{amsmath}
\usepackage{amssymb}
\usepackage{graphicx}
\graphicspath{ {./images/} }
\usepackage{booktabs}
\usepackage{url}
\usepackage{siunitx} 
\usepackage{xcolor}
\usepackage{lineno}
\usepackage{enumitem}
\usepackage[margin=2.5cm]{geometry}
\usepackage{tabularx}
\newcolumntype{R}{>{\raggedleft\arraybackslash}X}
\usepackage{multirow}
\usepackage{rotating}
\usepackage{subcaption}
\usepackage{tikz}
\usetikzlibrary{shapes.geometric, arrows, positioning, calc, shadows.blur, backgrounds, fit}
\usepackage{algorithm}
\usepackage{algpseudocode}

\makeatletter
\providecommand*{\toclevel@algorithm}{0}
\makeatother

\usepackage{makecell}
\usepackage[unicode]{hyperref} 

\journal{Electric Power Systems Research}

\definecolor{myblue}{HTML}{5499C7}
\definecolor{myorange}{HTML}{F39C12}
\definecolor{mygreen}{HTML}{52BE80}
\definecolor{myred}{HTML}{E74C3C}
\definecolor{mygray}{HTML}{EAECEE}

\tikzset{
	gnn_input/.style = {rectangle, rounded corners=3pt, minimum width=3.2cm, minimum height=0.9cm, text centered, draw=black, fill=myblue!80, text=white, align=center, blur shadow={shadow xshift=0.5ex, shadow yshift=-0.5ex}, font=\sffamily\footnotesize},
	gnn_layer/.style = {rectangle, rounded corners=3pt, minimum width=3.2cm, minimum height=1.2cm, text centered, draw=black, fill=myorange!80, text=white, align=center, blur shadow={shadow xshift=0.5ex, shadow yshift=-0.5ex}, font=\sffamily\footnotesize},
	gnn_output/.style = {rectangle, rounded corners=3pt, minimum width=3.2cm, minimum height=0.9cm, text centered, draw=black, fill=mygreen!80, text=white, align=center, blur shadow={shadow xshift=0.5ex, shadow yshift=-0.5ex}, font=\sffamily\footnotesize},
	gnn_arrow/.style = {thick,->,>= stealth, rounded corners=3pt},
	gnn_resid/.style = {circle, draw=black, fill=white, inner sep=0pt, minimum size=5mm, font=\sffamily\bfseries, blur shadow={shadow xshift=0.5ex, shadow yshift=-0.5ex}},
	io/.style = {trapezium, trapezium left angle=70, trapezium right angle=110, minimum width=3.2cm, minimum height=0.8cm, text centered, draw=black, fill=myblue!80, text=white, rounded corners=3pt, blur shadow={shadow xshift=0.5ex, shadow yshift=-0.5ex}, font=\sffamily\footnotesize},
	process/.style = {rectangle, minimum width=3.2cm, minimum height=0.8cm, text centered, text width=3.2cm, draw=black, fill=myorange!80, text=white, rounded corners=3pt, blur shadow={shadow xshift=0.5ex, shadow yshift=-0.5ex}, font=\sffamily\footnotesize},
	decision/.style = {diamond, aspect=2, minimum width=2.5cm, minimum height=0.8cm, text centered, draw=black, fill=myred!80, text=white, rounded corners=3pt, blur shadow={shadow xshift=0.5ex, shadow yshift=-0.5ex}, font=\sffamily\footnotesize},
	arrow/.style = {thick, ->, >= stealth, rounded corners=3pt},
	stage_box/.style = {draw=gray, thick, dashed, rounded corners=8pt, fill=mygray, fill opacity=0.3},
	train_box/.style = {draw=myorange, thick, dotted, rounded corners=8pt}
}

\makeatletter
\def\ps@pprintTitle{%
	\let\@oddhead\@empty
	\let\@evenhead\@empty
	\def\@oddfoot{}%
	\let\@evenfoot\@oddfoot}
\makeatother

\begin{document}
	
	\begin{frontmatter}
		
		\title{A Hybrid GNN-LSE Method for Fast, Robust, and Physically-Consistent AC Power Flow}
		\author[asu]{Mohamed Shamseldein}
		\address[asu]{Assistant Professor, Electrical Power and Machines Department, Ain Shams University, Cairo, Egypt}
		
		\begin{abstract}
			Conventional AC Power Flow (ACPF) solvers like Newton-Raphson (NR) face significant computational and convergence challenges in modern, large-scale power systems. This paper proposes a novel, two-stage hybrid method that integrates a Physics-Informed Graph Neural Network (GNN) with a robust, iterative Linear State Estimation (LSE) refinement step to produce fast and physically-consistent solutions. The GNN, trained with a physics-informed loss function featuring an efficient dynamic weighting scheme, rapidly predicts a high-quality initial system state. This prediction is then refined using an iterative, direct linear solver inspired by state estimation techniques. This LSE refinement step solves a series of linear equations to enforce physical laws, effectively bypassing the non-linearities and convergence issues of traditional solvers. The proposed GNN-LSE framework is comprehensively validated on systems ranging from small radial distribution networks (IEEE 33-bus, 69-bus) to a large, meshed transmission system (IEEE 118-bus). Results show that our GNN variants are up to \num[exponent-mode = scientific, round-mode=places, round-precision=1]{8.4e3} times faster than NR. The LSE refinement provides a fast route to a physically-consistent solution, while heavy-loading stress tests (\SI{120}{\percent}--\SI{150}{\percent} of nominal) and N-1 contingencies demonstrate the method's reliability and generalization. This work presents a powerful and flexible framework for bridging fast, data-driven models with the rigorous constraints of power system physics, offering a practical tool for real-time operations and analysis.
		\end{abstract}
		
		
	\end{frontmatter}
	
	\section*{Nomenclature and Acronyms}
	\begin{table}[!htbp]
		\footnotesize
		\begingroup\renewcommand\tabularxcolumn[1]{m{#1}}
		\begin{tabularx}{\textwidth}{>{\raggedright\arraybackslash}m{0.22\textwidth} >{\raggedright\arraybackslash}X}
			\toprule
			\textbf{Symbol/Acronym} & \textbf{Description} \\
			\midrule
			\multicolumn{2}{l}{\textbf{Acronyms}} \\
			ACPF & Alternating Current Power Flow \\
			FDPF & Fast Decoupled Power Flow \\
			GNN & Graph Neural Network \\
			KCL & Kirchhoff's Current Law \\
			LSE & Linear State Estimation \\
			MAE & Mean Absolute Error \\
			ML & Machine Learning \\
			MLP & Multi-Layer Perceptron \\
			MSE & Mean Squared Error \\
			NR & Newton-Raphson \\
			PINN & Physics-Informed Neural Network \\
			\midrule
			\multicolumn{2}{l}{\textbf{Variables and Parameters}} \\
			$|V_i|$, $\delta_i$ & Voltage magnitude and angle at bus $i$ \\
			$P_i$, $Q_i$ & Net active and reactive power injection at bus $i$ \\
			$P_{D,i}, Q_{D,i}$ & Active and reactive power demand at bus $i$ \\
			$P_{G,i}, V_{G,i}$ & Active power generation and specified voltage at bus $i$ \\
			$P_{S,i}$ & Real power injection at slack bus $i$ \\
			$Y_{\text{bus}}$ & Bus admittance matrix \\
			$G_{ik}$, $B_{ik}$ & Real and imaginary parts of the $(i,k)$-th element of $Y_{\text{bus}}$ \\
			$L_{\text{data}}$ & Data-driven loss component (MSE) \\
			$L_{\text{res}}$ & Physics-based residual loss component \\
			$L_{P}$, $L_{Q}$, $L_{V}$, $L_{S}$ & Active power, reactive power, voltage, and slack power residual losses \\
			$w_{PQ}$, $w_{V}$, $w_{S}$ & Dynamic weights for power, voltage, and slack residual losses \\
			$\beta$ & Momentum parameter for dynamic weight update \\
			$V^{(k)}$ & Voltage estimate at iteration $k$ of the LSE refinement \\
			\bottomrule
		\end{tabularx}\endgroup
	\end{table}
	
	\clearpage
	\section{Introduction}
	The Alternating Current Power Flow (ACPF) calculation is the bedrock of power system planning and operation, providing a snapshot of the grid's steady-state condition. The industry-standard Newton-Raphson (NR) method, while highly accurate, faces significant challenges with the increasing complexity of modern grids \cite{1, 2}. Its iterative, non-linear nature can lead to high computational costs and convergence failures, particularly in large-scale, heavily loaded, or ill-conditioned networks \cite{3}. These challenges are further amplified by the massive integration of uncertain loads, such as the widespread adoption of electric vehicles, which requires extensive and rapid contingency analysis to ensure grid stability \cite{4}. Faster variants like the Fast Decoupled Power Flow (FDPF) method offer a speed advantage by simplifying the Jacobian matrix, but their underlying assumptions (e.g., high X/R ratios) limit their reliability in distribution systems or networks with significant renewable penetration \cite{5, 6, 7}. This trade-off between speed, accuracy, and robustness has created a critical need for new ACPF solvers capable of meeting the demands of real-time operations.
	
	Machine Learning (ML), particularly Deep Learning, has emerged as a promising frontier for accelerating ACPF \cite{8, 9}. Early efforts with Multi-Layer Perceptrons (MLPs) showed potential but were fundamentally limited, as they treat the grid as a "black box" and fail to explicitly model the underlying topological structure, hindering their scalability and generalization \cite{10, 11}. Graph Neural Networks (GNNs) represent a significant leap forward, as they naturally model the power system as a graph of buses (nodes) and lines (edges). By using a message-passing mechanism, GNNs can learn the complex physical interdependencies dictated by the grid's topology \cite{12, 13, 14, 15}. However, a major barrier to their adoption in critical infrastructure is that standard GNNs, trained on data alone, provide no guarantee that their predictions will adhere to fundamental physical laws, such as Kirchhoff's laws \cite{16, 17, 18, 19}.
	
	To bridge this gap, Physics-Informed Neural Networks (PINNs) have gained traction. PINNs integrate physical principles directly into the training process by adding a physics-based loss term that penalizes deviations from governing equations \cite{11, 20, 21, 22, 23}. This approach not only promotes physically-consistent solutions but also acts as a powerful regularizer, often improving data efficiency and convergence \cite{24, 25, 26, 27, 28, 29, 30, 31}. Recent works have successfully combined GNNs with PINN concepts, creating models that are both topology-aware and physics-informed \cite{3, 32, 33}. However, many of these approaches still face challenges. For instance, the work in \cite{3} presents a topology-transferable GNN but relies on the GNN's output directly, which may still contain small but significant physical violations. Other methods often rely on complex and computationally intensive schemes for balancing data and physics losses during training.
	
	This paper introduces a novel, two-stage hybrid framework that leverages the speed of GNNs while enforcing physical consistency through a robust, linear post-processing step. Unlike prior works that stop at the GNN prediction, our key innovation is a second stage that refines the GNN's output using an iterative linear solver inspired by state estimation techniques. This LSE-style refinement step solves a series of partitioned linear equations to drive power mismatches toward zero. This approach elegantly sidesteps the convergence issues of non-linear solvers like NR while providing a fast and reliable path to a physically-consistent solution.
	
	The main contributions of this work are:
	\begin{itemize}[leftmargin=*, noitemsep]
		\item \textbf{A Novel Hybrid GNN-LSE Framework:} We propose a two-stage methodology that first uses a custom GNN-PINN for a rapid, high-quality ACPF prediction and then refines it with an iterative linear solver to enforce physical consistency.
		\item \textbf{Robust and Fast Linear Refinement:} We introduce the use of an iterative, partitioned linear solver as a fast alternative to non-linear refinement. This method is computationally efficient and robust against minor prediction errors from the GNN.
		\item \textbf{Comprehensive and Expanded Validation:} We rigorously evaluate our framework not only on standard radial distribution systems (IEEE 33- and 69-bus) but also demonstrate its scalability on a larger, meshed transmission network (IEEE 118-bus).
		\item \textbf{Extensive Robustness and Generalization Testing:} We validate the model's practical viability through a series of demanding tests, including heavy-loading stress (\SI{120}{\percent}--\SI{150}{\percent} of nominal) scenarios and N-1 contingencies, demonstrating its reliability beyond the training distribution. Under such beyond-training-distribution (heavy-loading) stress, enforcing power balance via LSE can increase angle MAE even as nodal power mismatches drop; this reflects physics-consistency being prioritized over pointwise agreement with the NR solution when the learned prior is far from the target.
	\end{itemize}
	
	\begin{sidewaystable}
		\centering
		\caption{Comparison of the Proposed Method with Recent Literature.}
		\label{tab:lit_review}
		\footnotesize
		\setlength{\tabcolsep}{3pt}
		\renewcommand{\arraystretch}{1.15}
		\begin{tabularx}{\textheight}{l c c c c c >{\raggedright\arraybackslash}X}
			\toprule
			\textbf{Study} & \textbf{ML Method} & \textbf{Physics-Informed} & \textbf{Refinement} & \textbf{Scalability Test} & \textbf{Robustness Test} & \textbf{Key Focus} \\
			\midrule
			Yang et al. \cite{3} & GNN & Yes (Loss) & No & Yes & No & Topology transferability for OPF \\
			Donon et al. \cite{18} & MLP & Yes (Loss) & No & No & No & Learning from physics without ground truth labels \\
			Varma et al. \cite{25} & MLP & Yes (Loss) & No & No & No & PINN for OPF \\
			Chen et al. \cite{29} & MLP & Yes (Regularizer) & Yes (Correction) & No & No & AC-OPF with feasibility guarantee \\
			Ngo et al. \cite{15} & GNN & Yes (Loss) & No & No & No & Physics-informed state estimation \\
			\midrule
			\textbf{This Work} & \textbf{GNN} & \textbf{Yes (Loss)} & \textbf{Yes (Iterative Linear)} & \textbf{Yes (118-bus)} & \textbf{Yes (Stress Test, N-1)} & \textbf{Fast ACPF with robust, linear refinement} \\
			\bottomrule
			\setlength{\tabcolsep}{6pt}
		\end{tabularx}
	\end{sidewaystable}
	
	\section{Methodology}
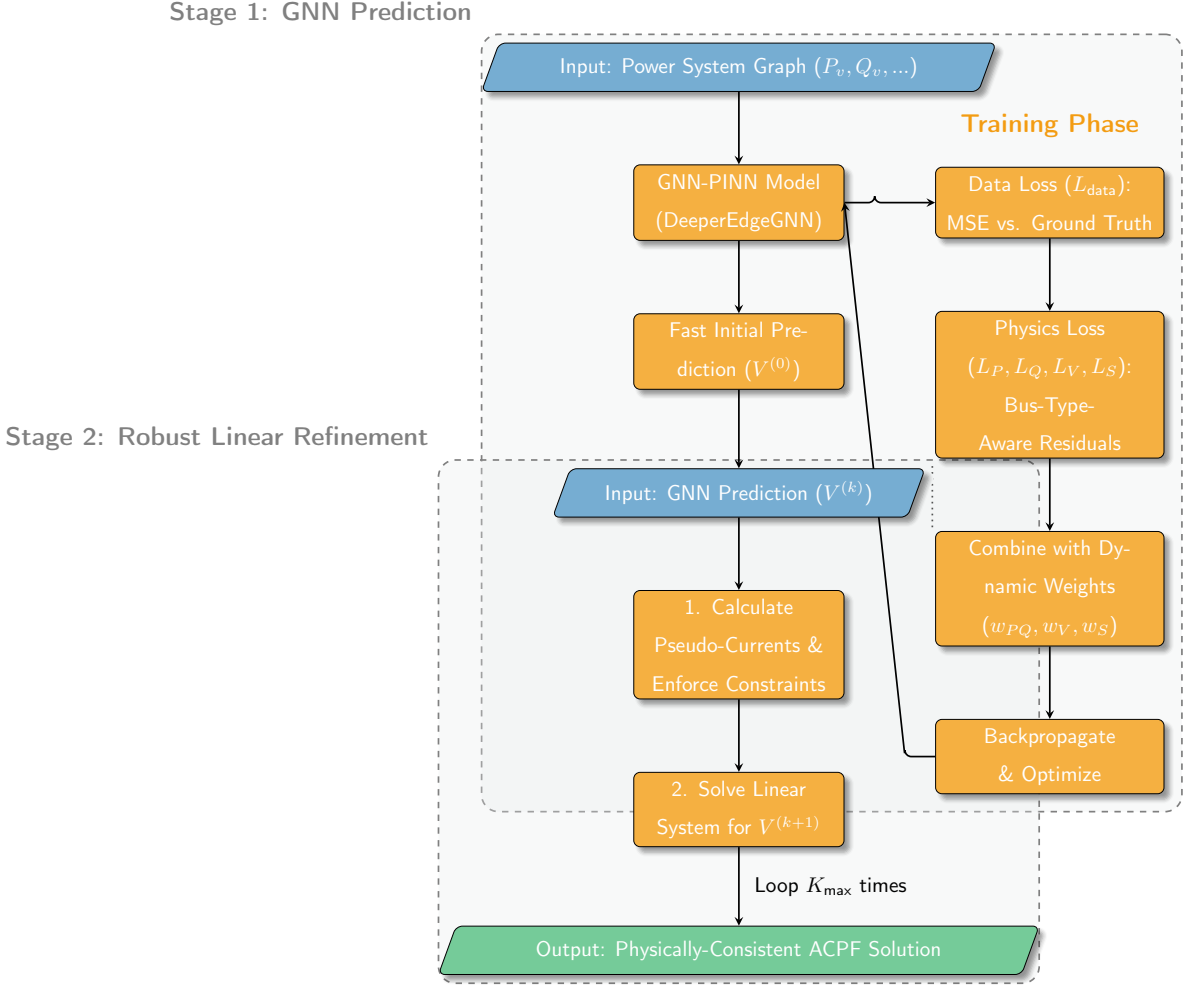
\begin{figure}[!htbp]
		\centering
		\resizebox{0.95\textwidth}{!}{%
			\begin{tikzpicture}[node distance=1.2cm and 0.4cm]
				\node (input) [io] {Input: Power System Graph ($P_v, Q_v, ...$)};
				\node (gnn) [process, below=of input] {GNN-PINN Model (DeeperEdgeGNN)};
				\node (predict) [process, below=of gnn] {Fast Initial Prediction ($V^{(0)}$)};
				
				\draw [arrow] (input) -- (gnn);
				\draw [arrow] (gnn) -- (predict);
				\node (loss_data) [process, right=1.5cm of gnn, text width=3.5cm] {Data Loss ($L_{\text{data}}$): MSE vs. Ground Truth};
				\node (loss_phys) [process, below=of loss_data, text width=3.5cm] {Physics Loss ($L_{P}, L_{Q}, L_{V}, L_{S}$): Bus-Type-Aware Residuals};
				\node (combine) [process, below=of loss_phys, text width=3.5cm] {Combine with Dynamic Weights ($w_{PQ}, w_{V}, w_{S}$)};
				\node (backprop) [process, below=of combine, text width=3.5cm] {Backpropagate \& Optimize};
				
				\draw [arrow] (gnn.east) -- ++(0.5,0) |- (loss_data);
				\draw [arrow] (loss_data) -- (loss_phys);
				\draw [arrow] (loss_phys) -- (combine);
				\draw [arrow] (combine) -- (backprop);
				\draw [arrow] (backprop.west) -| ++(-0.5,0) -- (gnn.east);
				
				\begin{scope}[on background layer]
					\node [train_box, fit=(loss_data) (backprop)] (train_fit) {};
				\end{scope}
				\node[above=2mm of train_fit, font=\sffamily\bfseries, text=myorange] {Training Phase};
				
				\node (refine_start) [io, below=1.3cm of predict] {Input: GNN Prediction ($V^{(k)}$)};
				\node (pseudo) [process, below=of refine_start] {1. Calculate Pseudo-Currents \& Enforce Constraints};
				\node (wls) [process, below=of pseudo] {2. Solve Linear System for $V^{(k+1)}$};
				\node (output) [io, below=1.3cm of wls, fill=mygreen!80] {Output: Physically-Consistent ACPF Solution};
				
				\draw [arrow] (predict) -- (refine_start);
				\draw [arrow] (refine_start) -- (pseudo);
				\draw [arrow] (pseudo) -- (wls);
				\draw [arrow] (wls.south) -- (output) node [midway, right=3pt, font=\sffamily\footnotesize] {Loop $K_{\text{max}}$ times};
				\begin{scope}[on background layer]
					\node [stage_box, fit=(input) (predict) (train_fit)] (stage1_fit) {};
					\node [stage_box, fit=(refine_start) (wls) (output)] (stage2_fit) {};
				\end{scope}
				\node[above left, font=\sffamily\bfseries, text=gray] at (stage1_fit.north west) {Stage 1: GNN Prediction};
				\node[above left, font=\sffamily\bfseries, text=gray] at (stage2_fit.north west) {Stage 2: Robust Linear Refinement};
			\end{tikzpicture}%
		}
		\caption{Flowchart of the proposed hybrid GNN-LSE methodology. Stage 1 involves training the GNN-PINN to produce a fast initial prediction. Stage 2 refines this prediction using a robust, iterative linear solver to yield a physically-consistent solution.}
		\label{fig:method_flowchart}
	\end{figure}

	This section details the proposed GNN-LSE methodology, describing the problem formulation, graph representation, GNN architecture, physics-informed loss function, and the refinement process. A flowchart summarizing the entire hybrid methodology is presented in Fig. \ref{fig:method_flowchart}.
	
	\subsection{Problem Formulation and Graph Representation}
	The ACPF problem aims to determine the voltage magnitudes ($|V|$) and phase angles ($\delta$) at all buses in a power grid for given load and generation conditions \cite{34, 35, 36}. The problem is defined by a set of non-linear algebraic equations describing the net active ($P_i$) and reactive ($Q_i$) power injections at each bus $i$:
	\begin{equation}
		P_i = \sum_{k=1}^{N} |V_i||V_k|(G_{ik} \cos(\delta_i - \delta_k) + B_{ik} \sin(\delta_i - \delta_k))
		\label{eq:p_inj}
	\end{equation}
	\begin{equation}
		Q_i = \sum_{k=1}^{N} |V_i||V_k|(G_{ik} \sin(\delta_i - \delta_k) - B_{ik} \cos(\delta_i - \delta_k))
		\label{eq:q_inj}
	\end{equation}
	where $G_{ik}$ and $B_{ik}$ are the real and imaginary parts of the bus admittance matrix ($Y_{\text{bus}}$).
	
	To leverage GNNs, the power system is modeled as a graph $G=(V, E)$, where buses are nodes and lines are edges \cite{37, 38}.
	\begin{itemize}[leftmargin=*]
		\item \textbf{Nodes (V):} Each bus is a node with five features $x_v$: its scaled net active power demand ($P_{D,v}$), reactive power demand ($Q_{D,v}$), active power generation ($P_{G,v}$), specified voltage magnitude ($V_{G,v}$), and an encoding for its bus type (PQ, PV, or slack).
		\item \textbf{Edges (E):} Each line is an edge with features $e_{uv}$ for its per-unit resistance ($R_{uv}$), reactance ($X_{uv}$), and half the total line charging susceptance ($B_{uv}/2$).
	\end{itemize}
	This graph structure serves as the direct input to the GNN model.
	
	\subsection{GNN Architecture}
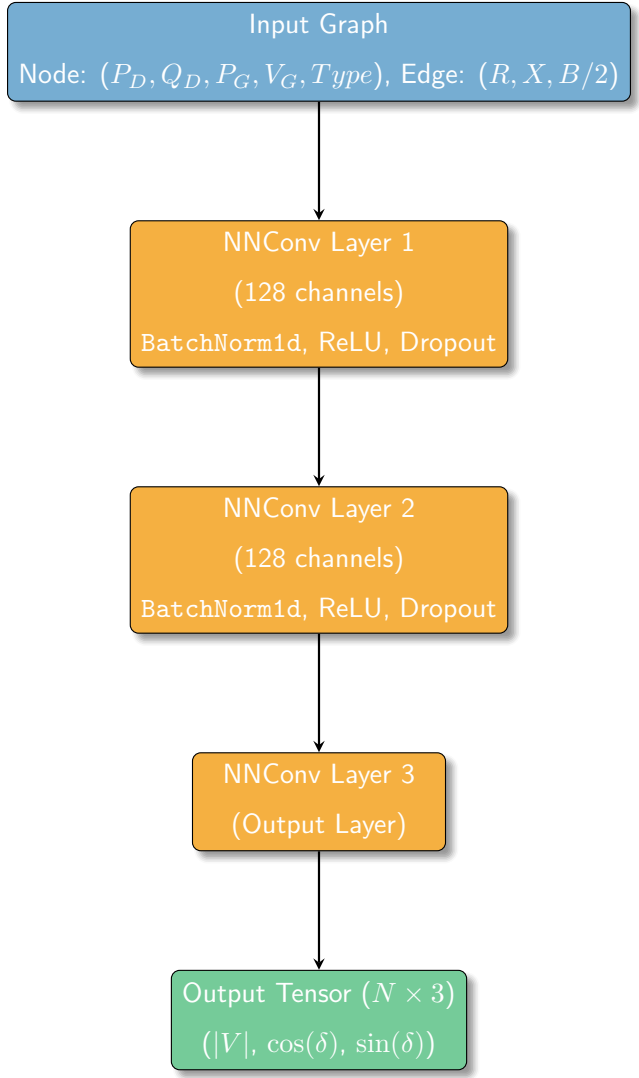
\begin{figure}[!htbp]
		\centering
		\resizebox{0.5\textwidth}{!}{%
			\begin{tikzpicture}[node distance=1.5cm and 1cm]
				\node (input) [gnn_input] {Input Graph \\ Node: ($P_D, Q_D, P_G, V_G, Type$), Edge: ($R, X, B/2$)};
				\node (layer1) [gnn_layer, below=of input] {NNConv Layer 1 \\ (128 channels) \\ \texttt{BatchNorm1d}, ReLU, Dropout};
				\node (layer2) [gnn_layer, below=of layer1] {NNConv Layer 2 \\ (128 channels) \\ \texttt{BatchNorm1d}, ReLU, Dropout};
				\node (layer3) [gnn_layer, below=of layer2] {NNConv Layer 3 \\ (Output Layer)};
				\node (output) [gnn_output, below=of layer3] {Output Tensor ($N \times 3$) \\ ($|V|$, $\cos(\delta)$, $\sin(\delta)$)};
				
				\draw [gnn_arrow] (input) -- (layer1);
				\draw [gnn_arrow] (layer1) -- (layer2);
				\draw [gnn_arrow] (layer2) -- (layer3);
				\draw [gnn_arrow] (layer3) -- (output);
			\end{tikzpicture}%
		}
		\caption{DeeperEdgeGNN architecture. The model uses three sequential NNConv layers without residual connections.}
		\label{fig:gnn_architecture}
	\end{figure}

	The GNN architecture, named DeeperEdgeGNN, maps the power system state to the ACPF solution using three sequential \texttt{torch\_geometric.nn.NNConv} layers from PyTorch Geometric \cite{38}. This layer type is particularly effective as it uses a small neural network to process edge features (line impedances), allowing it to learn complex interactions between connected buses \cite{39, 40}.
	\begin{itemize}[leftmargin=*]
		\item \textbf{Input}: The graph G with the five-dimensional node features and three-dimensional edge features.
		\item \textbf{Hidden Layers}: Two NNConv layers map node representations to a 128-dimension hidden space, each followed by \texttt{BatchNorm1d}, ReLU activation, and dropout (rate: 0.2).
		\item \textbf{Output Layer}: A final NNConv layer maps the hidden representations to three output features per bus: scaled voltage magnitude ($|\hat{V}_v|$), and the scaled cosine and sine of the voltage angle ($\cos(\hat{\delta}_v)$, $\sin(\hat{\delta}_v)$). Predicting sine and cosine provides a continuous representation that avoids angle wrapping issues.
	\end{itemize}
	The architecture, detailed in Fig. \ref{fig:gnn_architecture}, was designed to balance model capacity with computational cost for the target systems.
	
	\subsection{Physics-Informed Loss Function with Dynamic Weighting}
	A cornerstone of our methodology is the composite loss function, $L_{\text{total}}$, which combines a data-driven term with physics-based constraints. This approach is rooted in the concept of Physics-Informed Neural Networks (PINNs), which embed physical laws directly into the learning process by augmenting the loss function with a term that penalizes deviations from these laws.
	
	\subsubsection{Data-Driven Loss Component (\texorpdfstring{$L_{\text{data}}$}{L\_data})}
	This component ensures the model's predictions align with the ground-truth solutions. It is the Mean Squared Error (MSE) between the GNN's scaled predictions ($\hat{V}, \cos(\hat{\delta}), \sin(\hat{\delta})$) and the scaled ground-truth values ($V, \cos(\delta), \sin(\delta)$) for each sample $s$ and non-slack bus $i$. For clarity in this equation, all variables represent their scaled versions:
	\begin{equation}
		\begin{split}
			L_{\text{data}} = \frac{1}{N_{s}N_{b,ns}}\sum_{s,i \in \text{non-slack}} \Big( & (|\hat{V}_{s,i}| - |V_{s,i}|)^2 + (\cos(\hat{\delta}_{s,i}) - \cos(\delta_{s,i}))^2 \\
			&+ (\sin(\hat{\delta}_{s,i}) - \sin(\delta_{s,i}))^2 \Big)
		\end{split}
	\end{equation}
	
	\subsubsection{Physics-Based Residual Loss Component (\texorpdfstring{$L_{\text{res}}$}{L\_res})}
	This component enforces the physical laws of AC power flow by penalizing violations of the power balance and voltage setpoint equations. Our formulation is critically aware of the different bus types (PQ, PV, slack) and applies constraints accordingly. We employ the Huber loss for its robustness to the large residuals that can occur in early training stages.
	
	The Huber loss is a piecewise function that combines the best properties of MSE and Mean Absolute Error (MAE). It is defined as:
	\begin{equation}
		L_{\delta}(a) =
		\begin{cases}
			\frac{1}{2}a^2 & \text{for } |a| \le \delta \\
			\delta(|a| - \frac{1}{2}\delta) & \text{otherwise}
		\end{cases}
		\label{eq:huber}
	\end{equation}
	where $a$ is the error or residual, and $\delta$ is a threshold (see Appendix~A for details). For small errors, it behaves quadratically, providing strong gradients for fine-tuning. For large errors, it behaves linearly, preventing large physics violations from creating excessively large gradients that would destabilize the learning process.
	
	The physics loss is composed of four distinct residual terms, calculated after un-scaling the GNN's voltage predictions ($\hat{V}$):
	\begin{itemize}[leftmargin=*]
		\item \textbf{Active Power Residual Loss ($L_{P}$):} This loss is calculated for all non-slack buses (both PQ and PV) based on the mismatch between specified active power ($P_{i,\text{spec}}$) and calculated active power ($P_{i,\text{calc}}$) from Eq. \ref{eq:p_inj}.
		\item \textbf{Reactive Power Residual Loss ($L_{Q}$):} This loss is calculated \textit{only} for non-slack PQ buses, penalizing the mismatch between specified reactive power ($Q_{i,\text{spec}}$) and calculated reactive power ($Q_{i,\text{calc}}$) from Eq. \ref{eq:q_inj}.
		\item \textbf{Voltage Magnitude Residual Loss ($L_{V}$):} This loss is calculated \textit{only} for PV buses, penalizing the deviation of the predicted voltage magnitude ($|\hat{V}_i|$) from its specified setpoint ($V_{i,\text{spec}}$).
		\item \textbf{Slack Power Residual Loss ($L_{S}$):} A novel component that penalizes the mismatch between the ground-truth real power injection at the slack bus ($P_{S,i,\text{spec}}$, from the dataset) and the calculated injection ($P_{S,i,\text{calc}}$), which helps the model learn the system's overall power balance.
	\end{itemize}
	
	\subsubsection{Overall Loss and Dynamic Weighting}
	A key challenge in training PINNs is balancing the data and physics loss terms. We employ an efficient dynamic weighting scheme that updates three separate weights: $w_{PQ}$ for the power residuals, $w_{V}$ for the voltage residuals, and $w_S$ for the slack power residual. The total loss is:
	\begin{equation}
		L_{\text{total}} = w_{\text{data}}\frac{L_{\text{data}}}{s_{\text{data}}} + w_{PQ}\left(\frac{L_{P}}{s_{P}} + \frac{L_{Q}}{s_{Q}}\right) + w_{V}\frac{L_{V}}{s_{V}} + w_{S}\frac{L_{S}}{s_{S}}
		\label{eq:total_loss}
	\end{equation}
	where $s$ terms are initial scaling factors to normalize the magnitude of each loss component at the start of training. After each epoch, the physics weights are updated using an exponential moving average based on validation set performance. First, target weights are calculated as the ratio of the validation data loss to the corresponding validation physics losses. Then, the new weights are computed as:
	\begin{align}
		w_{\text{PQ}, t+1} &= (\beta \times w_{\text{PQ}, t}) + (1-\beta) \times \left( \frac{L_{\text{data, val}}}{L_{\text{P, val}} + L_{\text{Q, val}} + \epsilon} \right) \label{eq:weight_update_pq} \\
		w_{\text{V}, t+1} &= (\beta \times w_{\text{V}, t}) + (1-\beta) \times \left( \frac{L_{\text{data, val}}}{L_{\text{V, val}} + \epsilon} \right) \label{eq:weight_update_v} \\
		w_{\text{S}, t+1} &= (\beta \times w_{\text{S}, t}) + (1-\beta) \times \left( \frac{L_{\text{data, val}}}{L_{\text{S, val}} + \epsilon} \right) \label{eq:weight_update_s}
	\end{align}
	where $\beta$ is a momentum parameter and $\epsilon$ is a small constant to ensure numerical stability (see Appendix~A). This feedback-driven mechanism smoothly and independently adjusts the influence of each physics term relative to the data loss, promoting stable convergence without manual tuning.
	
	\subsection{Iterative and Robust Linear Refinement}
	While the GNN-PINN produces a fast and physically-plausible prediction, minor physical violations may persist. To mitigate these and obtain a physically-consistent final solution, we introduce a novel, iterative linear refinement step. This method treats the GNN's output voltage phasor, $V_{GNN}$, as a high-quality initial guess and refines it by solving a series of partitioned linear problems, avoiding the convergence issues of non-linear solvers.
	
	\subsubsection{Linearization and Problem Formulation}
	This refinement is based on a fixed-point iterative method, similar in principle to a partitioned Gauss-Seidel solver. The relationship between bus voltages ($V$) and current injections ($I$) is given by the linear system $I = Y_{\text{bus}}V$. The core challenge is that the injection currents for PQ buses depend non-linearly on their own unknown voltages ($I_i = (S_{i,\text{spec}}/V_i)^*$).
	
	Our iterative approach linearizes this problem by using the voltage estimate from the previous iteration, $V^{(k)}$, to calculate the currents. The bus admittance matrix is partitioned based on bus types (PQ, PV, and slack/reference `s`):
	\begin{equation}
		\begin{bmatrix} I_{pq} \\ I_{pv} \\ I_s \end{bmatrix} = \begin{bmatrix} Y_{pp} & Y_{pv} & Y_{ps} \\ Y_{vp} & Y_{vv} & Y_{vs} \\ Y_{sp} & Y_{sv} & Y_{ss} \end{bmatrix} \begin{bmatrix} V_{pq} \\ V_{pv} \\ V_s \end{bmatrix}
	\end{equation}
	For the PQ buses, the linear system to be solved at iteration $k+1$ becomes:
	\begin{equation}
		Y_{pp}V_{pq}^{(k+1)} = I_{pq}^{(k+1)} - Y_{pv}V_{pv}^{(k+1)} - Y_{ps}V_{s}
		\label{eq:lse_linear}
	\end{equation}
	where $V_s$ is fixed, the PV bus voltages $V_{pv}^{(k+1)}$ are updated by enforcing their specified magnitudes while using the angles from iteration $k$, and the injection currents for the PQ buses $I_{pq}^{(k+1)}$ are calculated using voltages from iteration $k$. This formulation allows for a direct, non-iterative solution for $V_{pq}^{(k+1)}$ at each step.
	
	\subsubsection{Refinement Algorithm}
	The complete iterative refinement process is detailed in Algorithm \ref{alg:lse_refinement}. The key steps are:
	\begin{enumerate}
		\item \textbf{Initialization:} The process starts with the GNN's prediction, $V^{(0)} = V_{GNN}$. To enhance robustness, initial voltage magnitudes are clipped to a physically plausible range [0.8, 1.2] p.u.
		\item \textbf{Iterative Refinement:} For a fixed number of iterations (typically 3), the algorithm performs the following:
		\begin{itemize}
			\item Updates PV bus voltage phasors by combining their specified magnitudes with the angles from the previous iteration's estimate.
			\item Calculates the required net current injections for all PQ buses ($I_{\text{net\_pq}}$) using the voltage estimates from the previous iteration.
			\item Solves the partitioned linear system (Eq. \ref{eq:lse_linear}) to get the new voltage phasors for all PQ buses, $V_{pq}^{(k+1)}$.
			\item Updates the full voltage vector with the new PQ voltages, the updated PV voltages, and the fixed slack voltage.
		\end{itemize}
	\end{enumerate}
	This iterative linear approach is highly efficient and robust. It is important to note that while this process significantly enhances physical consistency, it is a linear approximation and, unlike a full non-linear solver, does not guarantee a numerically exact solution to the original non-linear ACPF equations.
	
	\begin{algorithm}
		\caption{Iterative Linear Refinement}
		\label{alg:lse_refinement}
		\begin{algorithmic}[1]
			\State \textbf{Input:} GNN voltage prediction $V_{GNN}$, specified power $S_{\text{spec}}$, admittance matrix $Y_{\text{bus}}$
			\State \textbf{Parameters:} Max iterations $K_{\text{max}}$
			\State \textbf{Initialize:} $V^{(0)} \gets \text{clip}(|V_{GNN}|) \cdot e^{j\angle V_{GNN}}$, Partition $Y_{\text{bus}}$ into $Y_{pp}, Y_{pv}, Y_{ps}$
			\For{$k = 0$ to $K_{\text{max}}-1$}
			\State Update $V_{pv}^{(k+1)}$ using specified $|V|$ and $\angle V_{pv}^{(k)}$
			\State Calculate pseudo-currents $I_{\text{pq}}^{(k)} \gets (S_{\text{spec,pq}} / V_{pq}^{(k)})^*$
			\State Calculate injected currents $I_{\text{inj}} \gets Y_{pv}V_{pv}^{(k+1)} + Y_{ps}V_{s}$
			\State Calculate net current $I_{\text{net\_pq}} \gets I_{pq}^{(k)} - I_{\text{inj}}$
			\State $V_{pq}^{(k+1)} \gets \text{solve}(Y_{pp}, I_{\text{net\_pq}})$ \Comment{Solve linear system}
			\State Update full voltage vector $V^{(k+1)}$ with new $V_{pq}^{(k+1)}$, $V_{pv}^{(k+1)}$, and fixed $V_s$
			\EndFor
			\State \textbf{Return:} Refined voltage $V^{(K_{\text{max}})}$
		\end{algorithmic}
	\end{algorithm}
	
	\paragraph{Why the GNN warm-start helps.}
	Because the LSE step is local, initializing at $V^{(0)} \!=\! V_{\text{GNN}}$ places the iterate \emph{inside the basin of attraction} where the linearization is accurate.
	Evaluating the partitioned system near the true state yields a better-conditioned solve for $V_{pq}^{(k+1)}$, stabilizing updates on ill-conditioned cases.
	In addition, the GNN prediction obeys the paper's slack and PV conventions, reducing angle-reference and set-point mismatches.
	Together, these effects explain the near-zero divergence and 1--3 iteration convergence reported in Section~\ref{sec:results}.
	
	\paragraph{When can it hurt?}
	Under strong distribution shift, the refinement prioritizes power-balance feasibility; angle MAE can increase even as mismatches drop---reflecting physics-consistency over pointwise NR agreement.

	The GNN forward pass has a complexity of $\mathcal{O}(|\mathcal{E}| \times d^2)$ where $|\mathcal{E}|$ is the number of edges and $d$ is the hidden dimension. Each refinement iteration requires solving a linear system for the PQ buses, which has a complexity of $\mathcal{O}(N_{pq}^3)$ where $N_{pq}$ is the number of PQ buses. With $K_{\text{max}}=3$ iterations, the total refinement complexity is $\mathcal{O}(K_{\text{max}} \times N_{pq}^3)$. While this cubic scaling is a theoretical bottleneck, for the system sizes tested, the refinement step remains computationally very fast, as shown in the results.
	
	\subsection{Training and Evaluation Details}
	Input features ($P_D, Q_D, P_G, V_G$, BusType) and output targets ($|V|, \cos\delta, \sin\delta$) were scaled using `StandardScaler` to have zero mean and unit variance, which is crucial for training stability. For each test system, a dataset of \num{50000} operating conditions was generated by sampling net load demands at each bus from a uniform distribution between \SI{0}{\percent} and \SI{200}{\percent} of their nominal values, ensuring a wide range of realistic operating points. Ground truth solutions were obtained using the PYPOWER NR solver with a strict tolerance of \num{1e-8} \cite{35}. The data was split into training (70\%), validation (15\%), and test (15\%) sets. The model was trained to minimize $L_{\text{total}}$ using the AdamW optimizer. All training hyperparameters, including learning rate, batch size, and scheduler details, are provided in Appendix~A. Early stopping with a patience of 40 epochs was used to prevent overfitting.
	
	\section{Experimental Setup and Evaluation}
	The GNN-PINN's performance was benchmarked against: (1) the classical NR method; (2) the Fast Decoupled Power Flow (FDPF) method; (3) a data-only GNN ablation (GNN-DataOnly) trained without the physics loss; (4) an additional in-house ML baseline using a Multi-Layer Perceptron (MLP-PINN) to highlight the benefit of the GNN's topology-aware structure; (5) the GNN-PINN prediction refined by our robust linear solver; and (6) a full NR solve initialized by the GNN-PINN prediction. The MLP-PINN baseline consists of three hidden layers with 256 neurons each, ReLU activations, and is trained with the same physics-informed loss function as the GNN. Performance was assessed on prediction accuracy (MAE), physical consistency (power mismatch distribution), and computational time. Angle errors are computed after aligning the slack-bus angle reference between the candidate solution and the NR ground truth. The 'max mismatch' metric is defined as the maximum absolute power mismatch, $\max_i(|\Delta P_i|, |\Delta Q_i|)$, evaluated across all buses in the system for a given solution. All reported results are presented as mean $\pm$ standard deviation over 5 runs with different random seeds to ensure statistical robustness. The NR and FDPF benchmarks were run on the CPU, while all ML models (GNN and MLP variants) were trained and evaluated on the GPU to leverage parallel processing. Computational times for ML models were measured by averaging over 1000 inference runs to ensure stable measurements for ultra-fast operations; times reported as `{$<0.01$}` indicate a single inference time below the measurement resolution of \SI{0.01}{\milli\second}. All experiments were conducted in Python using PyTorch, PyTorch Geometric, and PYPOWER \cite{35} on an AMD Ryzen 7 CPU with an NVIDIA RTX 4070 GPU and \SI{40}{\giga\byte} RAM. Further details on the empirical validation of computational scaling are provided in Appendix~B.
	
	\section{Results and Discussion}
	\label{sec:results}

	\subsection{Performance on Benchmark Distribution Systems}
\begin{figure*}[!htbp]
		\centering
		\begin{subfigure}[b]{0.8\textwidth}
			\includegraphics[width=\textwidth]{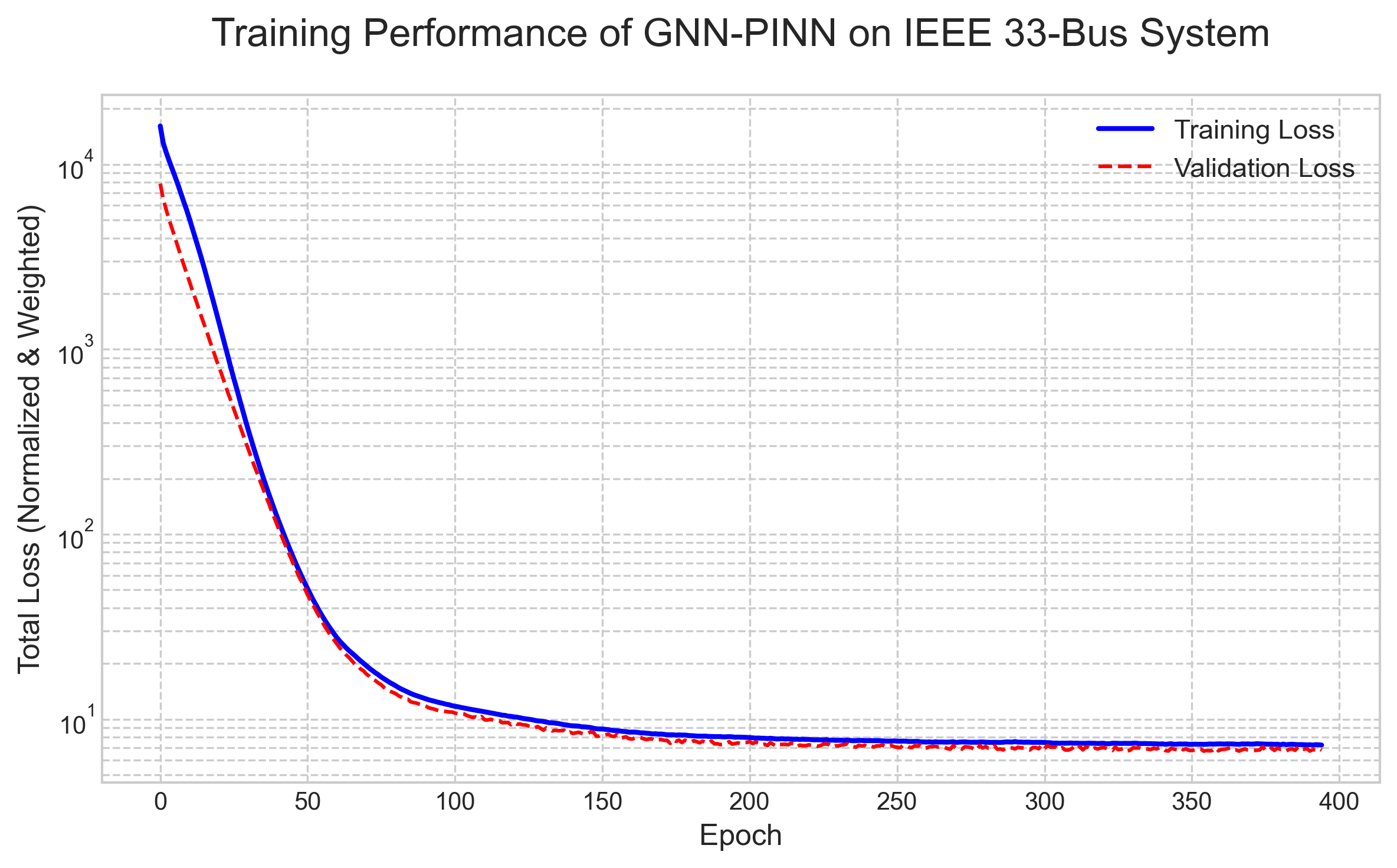}
			\caption{GNN-PINN Total Loss (33-Bus)}
		\end{subfigure}
		\hfill
		\begin{subfigure}[b]{0.8\textwidth}
			\includegraphics[width=\textwidth]{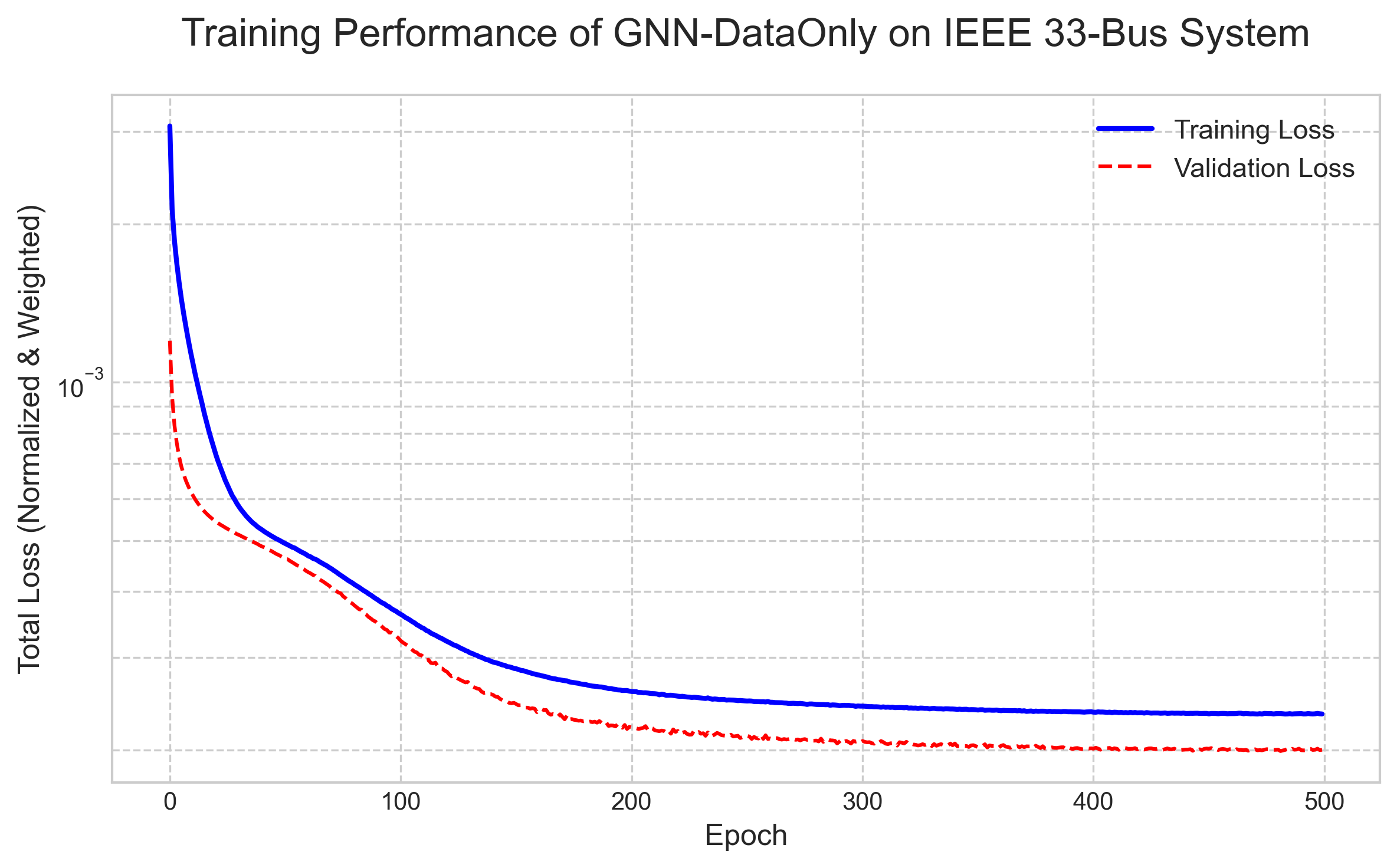}
			\caption{GNN-DataOnly Total Loss (33-Bus)}
		\end{subfigure}
		\caption{Training performance on the IEEE 33-bus system.}
		\label{fig:training_plots}
	\end{figure*}

	On the IEEE 33-bus system, the GNN-PINN demonstrated superior training efficiency compared to the GNN-DataOnly model (Fig. \ref{fig:training_plots}). The physics-informed loss acts as a powerful regularizer, guiding the model toward a physically valid solution space more directly.
	
	For the IEEE-69 test set (7{,}500 cases), only 5 scenarios (0.067\%) exceeded 100 p.u. mismatch in the \emph{raw} GNN output. With $K_{\max}=3$, all 7{,}500 cases---including those five---successfully refined to $\max(|\Delta P|,|\Delta Q|) \le 10^{-8}$ p.u.
	
	The refinement methods offer a clear choice between speed and precision. The GNN-PINN alone is thousands of times faster than a full NR solution. The robust linear solver (+LSE) provides an excellent balance, running many times faster than NR while significantly improving physical consistency. In contrast, refining with 5 NR iterations on the 69-bus system, while driving power mismatches to near-zero, can converge to a physically incorrect state, yielding a large average angle error of \SI{7.86}{\degree} (Table \ref{tab:results_69bus}) and highlighting the stability advantages of our direct linear refinement method. It is important to note, however, that the GNN provides a high-quality starting point, which can improve the convergence reliability of NR, especially in stressed systems where a flat start might fail.
	
	\begin{table}[!htbp]
		\setlength{\tabcolsep}{3pt}\renewcommand{\arraystretch}{1.1}
		\centering
		\caption[Performance and Accuracy on the IEEE 33-Bus System]{Performance and Accuracy on the IEEE 33-Bus System (Mean $\pm$ Std. Dev. over 5 runs).}
		\label{tab:results_33bus}
		\footnotesize
		\begin{tabular*}{\linewidth}{@{\extracolsep{\fill}} l S[table-format=1.1e-3, separate-uncertainty] S[table-format=2.2, separate-uncertainty] S[table-format=2.2, separate-uncertainty] c S[table-format=2.2, separate-uncertainty] S[table-format=2.2, separate-uncertainty]}
			\toprule
			\multirow{2}{*}{\textbf{Method}} & \multicolumn{2}{c}{\makecell{\textbf{Accuracy}\\\textbf{(MAE)}}} & \multicolumn{2}{c}{\makecell{\textbf{Performance}}} & \multicolumn{2}{c}{\makecell{\textbf{Phys. Consistency}\\\textbf{(max mismatch, p.u.)}}} \\
			\cmidrule(lr){2-3} \cmidrule(lr){4-5} \cmidrule(lr){6-7}
			& {\textbf{$|V_m|$ (p.u.)}} & {\textbf{$V_a$ (deg)}} & {\textbf{Time (ms)}} & {\textbf{ ($\times$NR)}} & {\textbf{$\Delta P$}} & {\textbf{$\Delta Q$}} \\
			\midrule
			GNN-PINN   & \num{0.003 +- 0.001} & \num{0.10 +- 0.05}  & {$<0.01$} & {$\ge \num[round-precision=2]{6.09e3}$} & \num{39.62 +- 31.42} & \num{34.89 +- 25.71} \\
			GNN (Data) & \num{0.003 +- 0.001} & \num{0.10 +- 0.05}  & {$<0.01$} & {$\ge \num[round-precision=2]{7.83e3}$} & \num{32.70 +- 25.46} & \num{37.19 +- 30.95} \\
			MLP-PINN & \num{0.001 +- 0.001} & \num{0.01 +- 0.01} & {$<0.01$} & {$\ge \num[round-precision=2]{7.92e3}$} & \num{32.69 +- 23.74} & \num{24.34 +- 15.51} \\
			+LSE       & \num{0.000 +- 0.000} & \num{0.00 +- 0.00}  & \num{1.52 +- 0.33} & \num{7.5}    & {$<10^{-8}$} & {$<10^{-8}$} \\
			+5NR & \num{0.000 +- 0.000} & \num{12.44 +- 7.99} & \num{13.49 +- 2.97} & \num{0.8}    & {$<10^{-8}$} & {$<10^{-8}$} \\
			NR         & \num{0.000 +- 0.000} & \num{0.00 +- 0.00}  & \num{11.42 +- 2.46} & \num{1.0}    & {$<10^{-8}$} & {$<10^{-8}$} \\
			FDPF       & \num{0.000 +- 0.000} & \num{0.00 +- 0.00}  & \num{6.94 +- 1.39} & \num{1.6}     & {$<10^{-8}$} & {$<10^{-8}$} \\
			\bottomrule
		\end{tabular*}
		
		\begin{minipage}{\textwidth}
			\footnotesize\raggedright
			\textit{Note: Mismatches are in p.u. on the system's standard base (\SI{10}{\mega\volt\ampere}). ``+5NR'' refers to five NR iterations (not run to tolerance) initialized from the GNN prediction. ``NR'' refers to a full solve from a flat start.}
		\end{minipage}
		
	\end{table}
	\setlength{\tabcolsep}{6pt}\renewcommand{\arraystretch}{1.0}
	
	\begin{table}[!htbp]
		\setlength{\tabcolsep}{3pt}\renewcommand{\arraystretch}{1.1} 
		\centering
		\caption[Performance and Accuracy on the IEEE 69-Bus System]{Performance and Accuracy on the IEEE 69-Bus System (Mean $\pm$ Std. Dev. over 5 runs).}
		\label{tab:results_69bus}
		\footnotesize
		\begin{tabular*}{\textwidth}{@{\extracolsep{\fill}} l S[table-format=1.1e-3, separate-uncertainty] S[table-format=1.2, separate-uncertainty] S[table-format=2.2] c S[table-format=4.2, separate-uncertainty] S[table-format=4.2, separate-uncertainty]}
			\toprule
			\multirow{2}{*}{\textbf{Method}} & \multicolumn{2}{c}{\makecell{\textbf{Accuracy}\\\textbf{(MAE)}}} & \multicolumn{2}{c}{\makecell{\textbf{Performance}}} & \multicolumn{2}{c}{\makecell{\textbf{Phys. Consistency}\\\textbf{(max mismatch, p.u.)}}} \\
			\cmidrule(lr){2-3} \cmidrule(lr){4-5} \cmidrule(lr){6-7}
			& {\textbf{$|V_m|$ (p.u.)}} & {\textbf{$V_a$ (deg)}} & {\textbf{Time (ms)}} & {\textbf{ ($\times$NR)}} & {\textbf{$\Delta P$}} & {\textbf{$\Delta Q$}} \\
			\midrule
			GNN-PINN   & \num{0.003 +- 0.001} & \num{2.55 +- 5.37} & {$<0.01$} & {$\ge \num[round-precision=2]{6.14e3}$} & \num{5703.85 +- 8255.62} & \num{3200.25 +- 3284.87} \\
			GNN (Data) & \num{0.003 +- 0.001} & \num{2.56 +- 5.37} & {$<0.01$} & {$\ge \num[round-precision=2]{7.25e3}$} & \num{5850.47 +- 8017.89} & \num{2961.14 +- 3132.90} \\
			MLP-PINN & \num{0.001 +- 0.001} & \num{6.44 +- 8.37} & {$<0.01$} & {$\ge \num[round-precision=2]{7.94e3}$} & \num{7029.46 +- 9344.52} & \num{4186.81 +- 4561.99} \\
			+LSE       & \num{0.000 +- 0.000} & \num{0.00 +- 0.00} & \num{1.74} & \num{6.8}    & {$<10^{-8}$} & {$<10^{-8}$} \\
			+5NR & \num{0.000 +- 0.000} & \num{7.86 +- 6.38} & \num{14.60} & \num{0.8}    & {$<10^{-8}$} & {$<10^{-8}$} \\
			NR   & \num{0.000 +- 0.000} & \num{0.00 +- 0.00} & \num{11.80} & \num{1.0}    & {$<10^{-8}$} & {$<10^{-8}$} \\
			FDPF       & \num{0.000 +- 0.000} & \num{0.00 +- 0.00} & \num{7.58} & \num{1.6}     & {$<10^{-8}$} & {$<10^{-8}$} \\
			\bottomrule
		\end{tabular*}
		
		\begin{minipage}{\textwidth}
			\footnotesize\raggedright
			\textit{Note: Mismatches are in p.u. on the system's standard base (\SI{10}{\mega\volt\ampere}). ``+5NR'' refers to five NR iterations (not run to tolerance) initialized from the GNN prediction. ``NR'' refers to a full solve from a flat start.}
		\end{minipage}
		
	\end{table}
	\setlength{\tabcolsep}{6pt}\renewcommand{\arraystretch}{1.0}

	\begin{figure*}[!htbp]
		\centering
		\begin{subfigure}[b]{0.8\textwidth}
			\includegraphics[width=\textwidth, keepaspectratio]{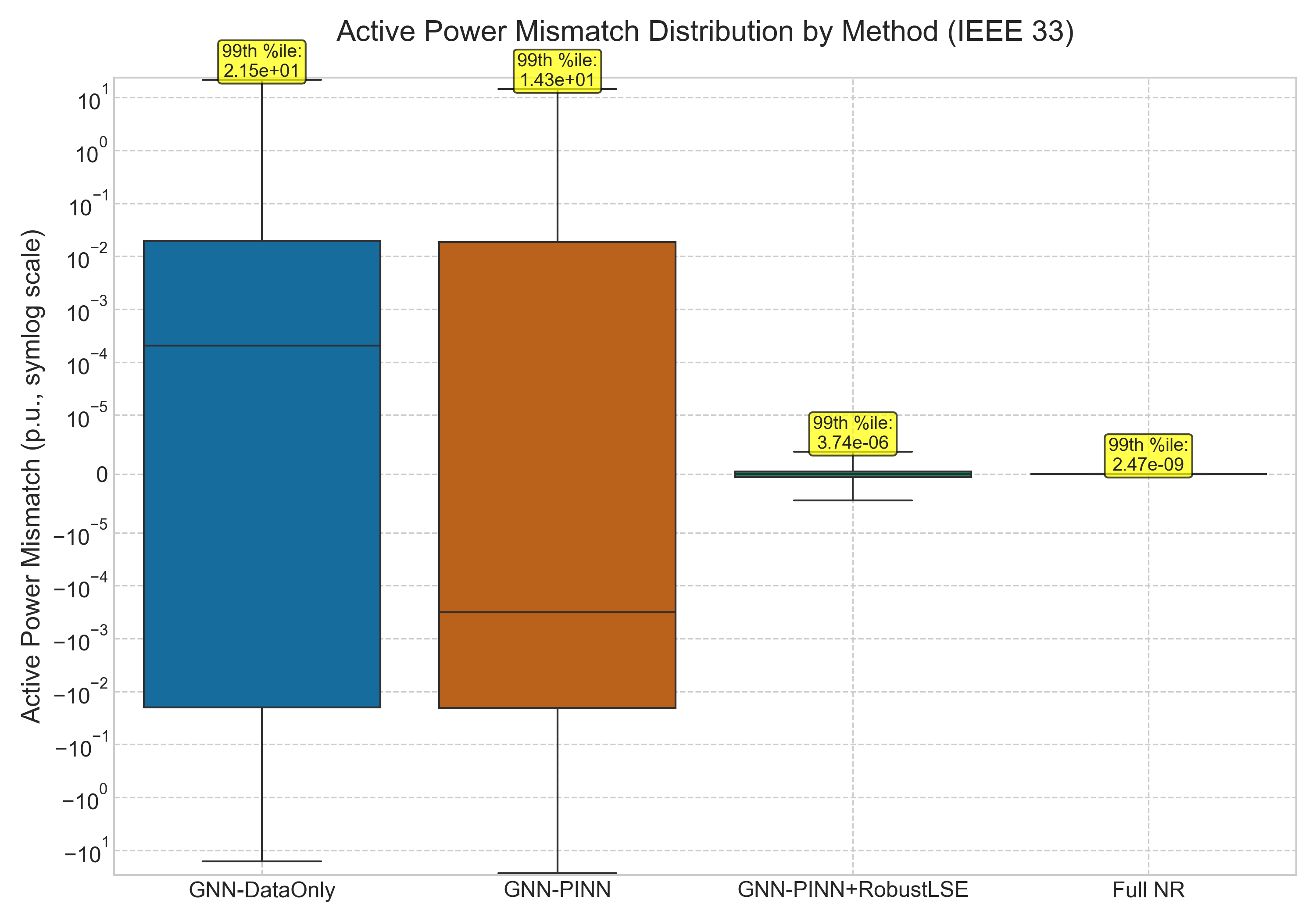}
			\caption[Active Power Mismatch (Delta P) (33-Bus)]{Active Power Mismatch ($\Delta P$) (33-Bus)}
		\end{subfigure}
		\hfill
		\begin{subfigure}[b]{0.8\textwidth}
			\includegraphics[width=\textwidth, keepaspectratio]{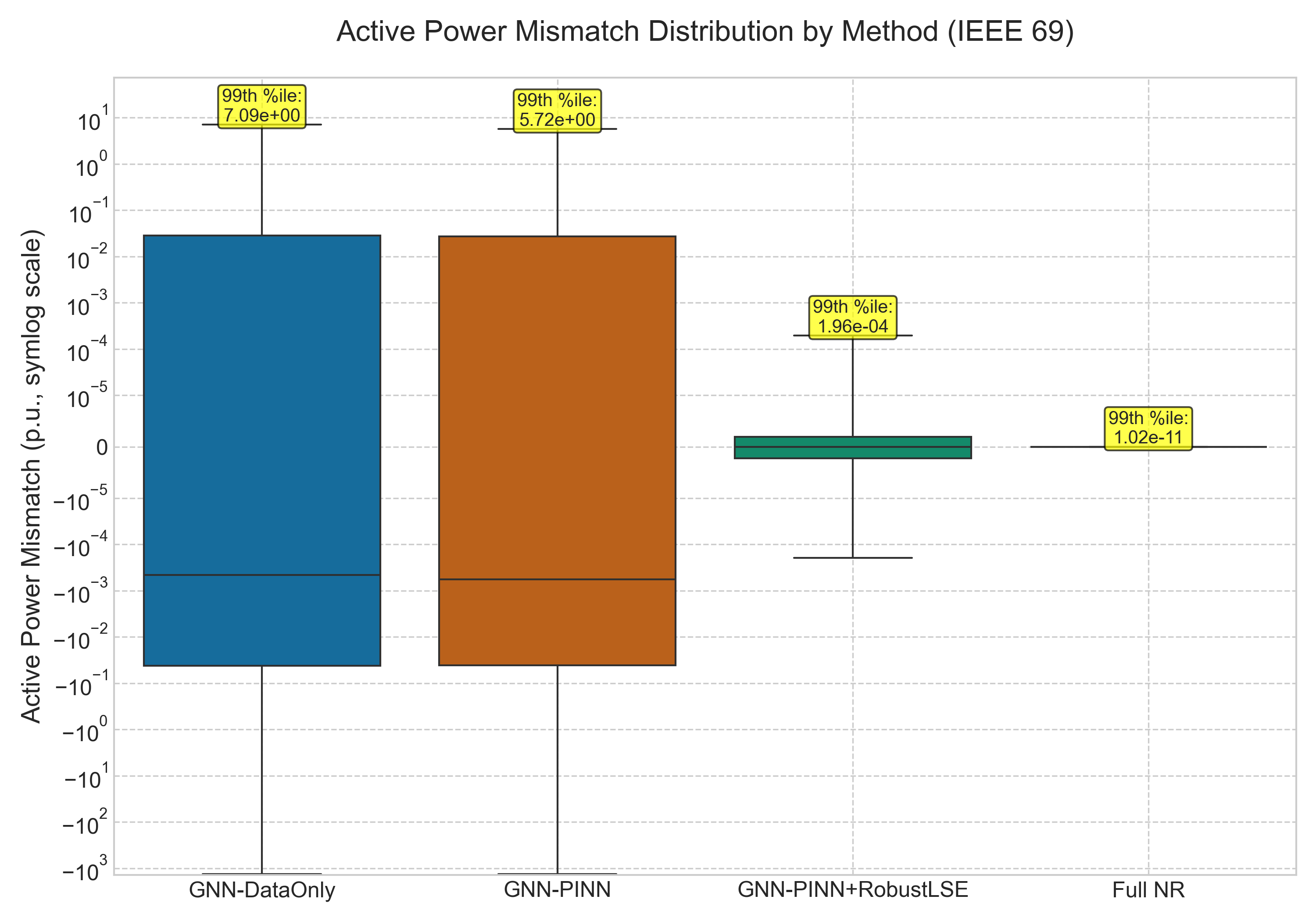}
			\caption[Active Power Mismatch (Delta P) (69-Bus)]{Active Power Mismatch ($\Delta P$) (69-Bus)}
		\end{subfigure}
		\caption{Distribution of active power mismatches across the distribution test systems. The boxplots clearly show the significant reduction in both median error and outlier magnitude achieved by the PINN approach and LSE refinement compared to the data-only model.}
		\label{fig:combined_results_all}
	\end{figure*}
	
	\subsection{Scalability on a Large, Meshed Transmission System}
\begin{figure*}[!htbp]
		\centering
		\includegraphics[width=0.8\textwidth, keepaspectratio]{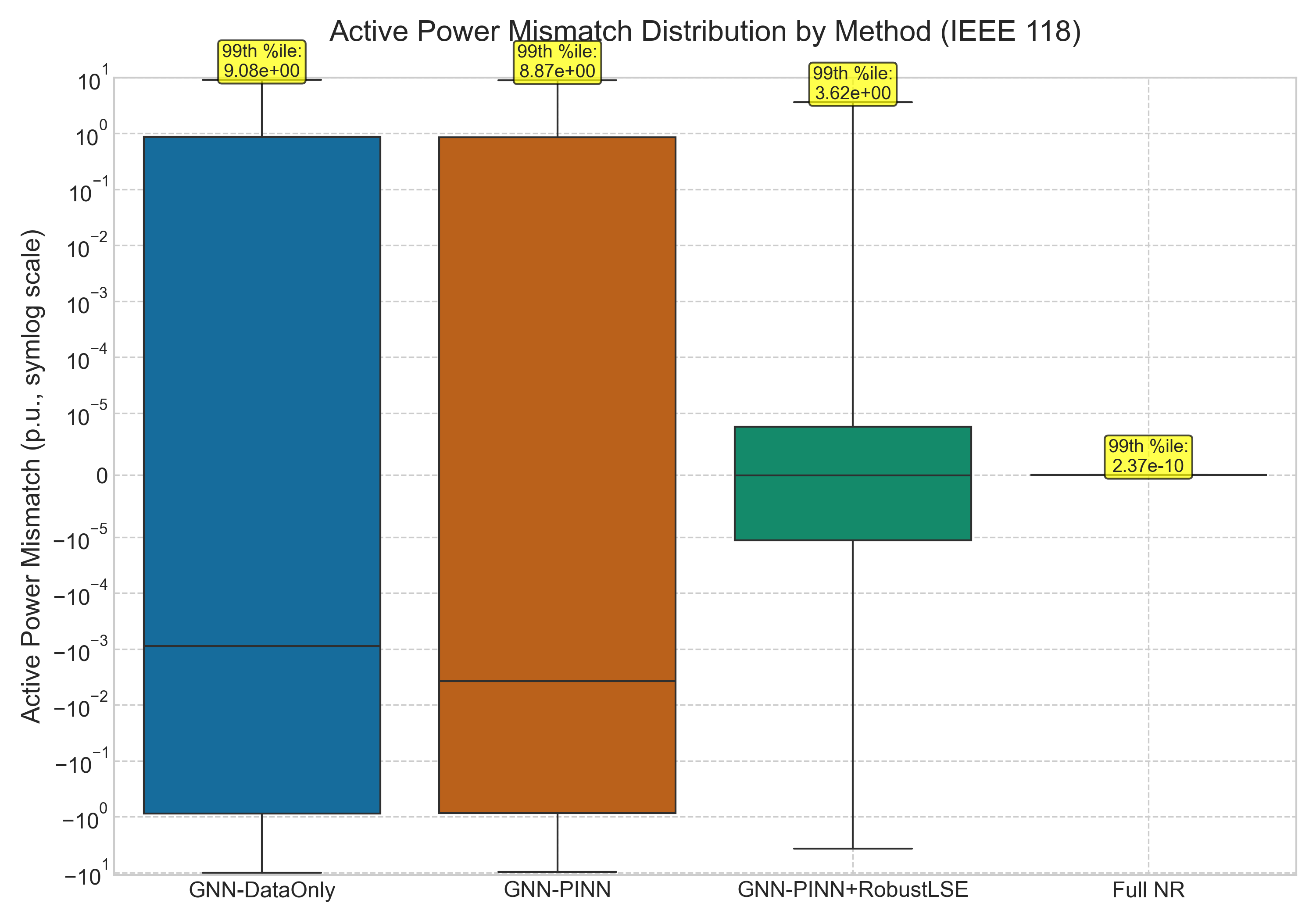}
		\caption{Distribution of active power mismatches for the IEEE 118-bus system.}
		\label{fig:results_118bus_boxplot}
	\end{figure*}

	To address the critical question of scalability, we evaluated our framework on the IEEE 118-bus system, a benchmark for larger, meshed transmission networks. As shown in Table \ref{tab:results_118bus}, the proposed method maintains its advantages. The GNN-PINN achieves a speedup of over \num{6900}$\times$ compared to the NR solver. The linear refinement step remains highly effective, reducing the maximum power mismatch of the GNN's prediction by over 63\% with only a minor computational cost, as shown in the mismatch distribution in Fig. \ref{fig:results_118bus_boxplot}. This demonstrates that the GNN can learn the more complex dependencies of a meshed grid and that the linear refinement remains a valid and efficient method for enforcing physical consistency at a larger scale.
	
	\begin{table}[!htbp]
		\setlength{\tabcolsep}{3pt}\renewcommand{\arraystretch}{1.1}
		\centering
		\caption[Performance and Accuracy on the IEEE 118-Bus System]{Performance and Accuracy on the IEEE 118-Bus System (Mean $\pm$ Std. Dev. over 5 runs).}
		\label{tab:results_118bus}
		\footnotesize
		\begin{tabular*}{\textwidth}{@{\extracolsep{\fill}} l S[table-format=1.2e-2, separate-uncertainty] S[table-format=2.2, separate-uncertainty] S[table-format=2.2] c S[table-format=2.2, separate-uncertainty] S[table-format=2.2, separate-uncertainty]}
			\toprule
			\multirow{2}{*}{\textbf{Method}} & \multicolumn{2}{c}{\makecell{\textbf{Accuracy}\\\textbf{(MAE)}}} & \multicolumn{2}{c}{\makecell{\textbf{Performance}}} & \multicolumn{2}{c}{\makecell{\textbf{Phys. Consistency}\\\textbf{(max mismatch, p.u.)}}} \\
			\cmidrule(lr){2-3} \cmidrule(lr){4-5} \cmidrule(lr){6-7}
			& {\textbf{$|V_m|$ (p.u.)}} & {\textbf{$V_a$ (deg)}} & {\textbf{Time (ms)}} & {\textbf{ ($\times$NR)}} & {\textbf{$\Delta P$}} & {\textbf{$\Delta Q$}} \\
			\midrule
			GNN-PINN   & \num{0.035 +- 0.007} & \num{4.98 +- 1.88}   & {$<0.01$}  & {$\ge \num[round-precision=2]{6.96e3}$} & \num{20.68 +- 13.99} & \num{15.10 +- 5.88} \\
			GNN (Data) & \num{0.035 +- 0.007} & \num{4.99 +- 1.88}   & {$<0.01$}  & {$\ge \num[round-precision=2]{8.40e3}$} & \num{20.54 +- 11.91} & \num{15.25 +- 5.98} \\
			MLP-PINN & \num{0.001 +- 0.001} & \num{0.65 +- 0.14} & {$<0.01$} & {$\ge \num[round-precision=2]{9.24e3}$} & \num{15.45 +- 6.47} & \num{9.84 +- 7.49} \\
			+LSE       & \num{0.001 +- 0.001} & \num{4.72 +- 1.89}   & \num{3.58}  & \num{4.0}    & \num{5.48 +- 2.03}   & \num{0.00 +- 0.01} \\
			+5NR & \num{0.000 +- 0.000} & \num{24.82 +- 15.07} & \num{17.60} & \num{0.8}    & {$<10^{-8}$} & {$<10^{-8}$} \\
			NR   & \num{0.000 +- 0.000} & \num{0.00 +- 0.00}  & \num{14.40} & \num{1.0}    & {$<10^{-8}$} & {$<10^{-8}$} \\
			FDPF       & \num{0.000 +- 0.000} & \num{0.00 +- 0.00}  & \num{9.16} & \num{1.6}     & {$<10^{-8}$} & {$<10^{-8}$} \\
			\bottomrule
		\end{tabular*}
		
		\begin{minipage}{\textwidth}
			\footnotesize\raggedright
			\textit{Note: Mismatches are in p.u. on the system's standard base (\SI{100}{\mega\volt\ampere}). ``+5NR'' refers to five NR iterations (not run to tolerance) initialized from the GNN prediction. ``NR'' refers to a full solve from a flat start.}
		\end{minipage}
		
	\end{table}
	\setlength{\tabcolsep}{6pt}\renewcommand{\arraystretch}{1.0}

	\subsection{Analysis of GNN Prediction Outliers}
	\label{sec:outlier_analysis}
	The maximum mismatch values reported for the raw GNN and MLP outputs in Table \ref{tab:results_69bus} warrant specific attention. These large values are reported on the system's \SI{10}{\mega\volt\ampere} base, a standard convention for this particular distribution test case, whereas the \SI{118}-bus transmission system uses a \SI{100}{\mega\volt\ampere} base. The enormous per-unit magnitudes confirm that they are numerical artifacts caused by a small number of catastrophic prediction outliers. It is critical to remember that the calculated power is a non-linear function of voltage, with terms proportional to the product of voltage magnitudes ($|V_i||V_k|$). Consequently, even a localized error in a single bus voltage prediction can be magnified, leading to the massive power mismatch values observed in these rare outlier cases. These outliers typically occur when the input represents an extreme loading condition at the edge of the training distribution. For the IEEE 69-bus system, out of 7,500 test scenarios, a maximum power mismatch exceeding 100 p.u. was observed in only 5 cases (0.067\%). Across all 7,500 69-bus test cases with $K_{\max}=3$, we observed $\mathbf{0}$ failed refinements; the five catastrophic outliers (0.067\%) all successfully converged to $\max(|\Delta P|,|\Delta Q|)\le 10^{-8}$ p.u. after refinement.
	
\begin{figure*}[!htbp]
		\centering
		\includegraphics[width=\textwidth]{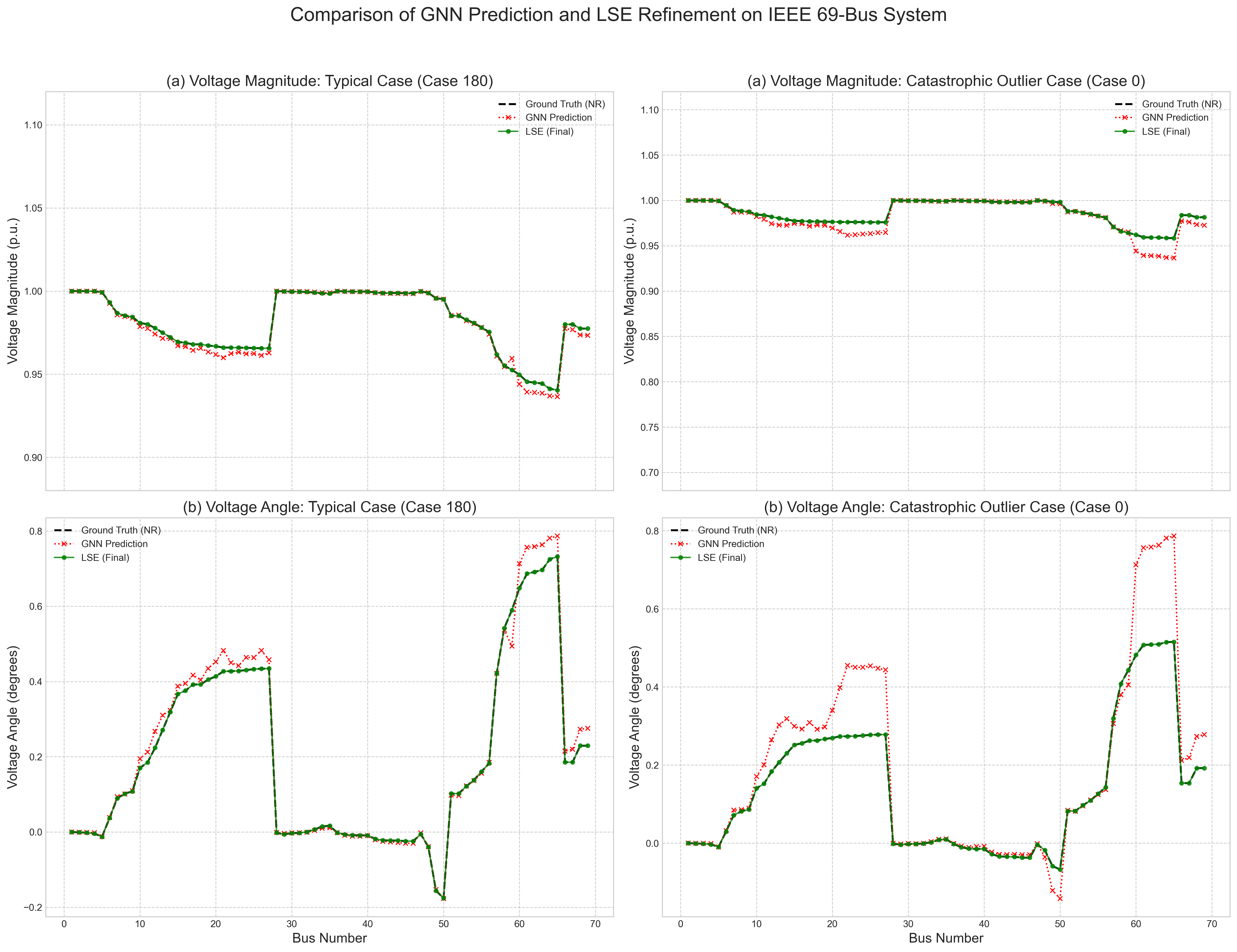}
		\caption{Comparison of GNN-PINN prediction and LSE refinement for a typical case (left) versus a catastrophic outlier case (right) on the IEEE 69-bus system. In the typical case, the GNN provides a high-quality initial estimate that is slightly improved by the LSE step. In the rare outlier case, the GNN exhibits a localized, brittle failure, which is robustly corrected by the LSE refinement, demonstrating the efficacy of the hybrid approach.}
		\label{fig:combined_profile_analysis}
	\end{figure*}

	The existence of these outliers highlights a key challenge: a purely data-driven model can exhibit \textit{brittle failure}. When presented with an extreme input, its output can become physically nonsensical without warning, as visualized in the erroneous voltage profiles of the catastrophic outlier case (right panels) in Figure \ref{fig:combined_profile_analysis}. While the physics-informed loss acts as a regularizer to mitigate this, our LSE refinement stage provides a true \textit{graceful failure} mechanism. It acts as a safety net that catches these rare, brittle failures and, as shown by the convergence traces in Figure \ref{fig:convergence_traces}, rapidly pulls the solution back into a physically valid state. This powerfully validates the necessity and robustness of the hybrid approach, demonstrating its ability to correct even poor initializations and ensure a reliable final output.

	\subsection{Robustness and Generalization Analysis}
	To assess the practical reliability of our method, we conducted a series of robustness tests beyond the standard i.i.d. test set.
	
	\subsubsection{\texorpdfstring{Heavy-Loading Stress Test (120--150\% of nominal)}{Heavy-Loading Stress Test (120--150\% of nominal)}}
	We evaluated the model, trained on loads from 0--200\%, on a new test set with heavy loading conditions (120--150\% of nominal). As shown in Tables \ref{tab:robustness_results_118} and \ref{tab:robustness_results_69}, while the prediction error (MAE) increases as expected, the GNN-PINN remains stable. The linear refinement step continues to be effective in reducing mismatches, demonstrating the model's ability to generalize to unseen, stressed operating conditions. As noted in Section 2.4, this can result in a higher angle MAE post-refinement, which reflects the LSE prioritizing physical power balance over pointwise agreement with a potentially unphysical NR solution under extreme stress.
	
	\subsubsection{Performance under N-1 Contingencies}
	We tested the model's ability to handle topological changes by evaluating it on 100 random N-1 line contingencies in the IEEE 118-bus system. The GNN's graph-based structure allows it to adapt to the modified topology without retraining. The results (Table \ref{tab:robustness_results_118}) show a graceful degradation in performance, with the GNN-LSE framework still providing fast and reasonably accurate solutions, highlighting its potential for real-time contingency analysis.
	
	\subsubsection{LSE Resilience to Prediction Errors}
	To validate the robustness of the linear refinement, we examined its behavior when initialized with GNN predictions that contained errors. The heavy-loading stress and N-1 tests show that even when the initial GNN error increases, the iterative linear solver remains stable and effectively reduces power mismatches, converging to a physically consistent state. This confirms the robustness of the refinement mechanism against imperfect initial predictions.

	\subsubsection{Convergence Behavior of LSE Refinement}
\begin{figure*}[!htbp]
		\centering
		\includegraphics[width=\textwidth, keepaspectratio]{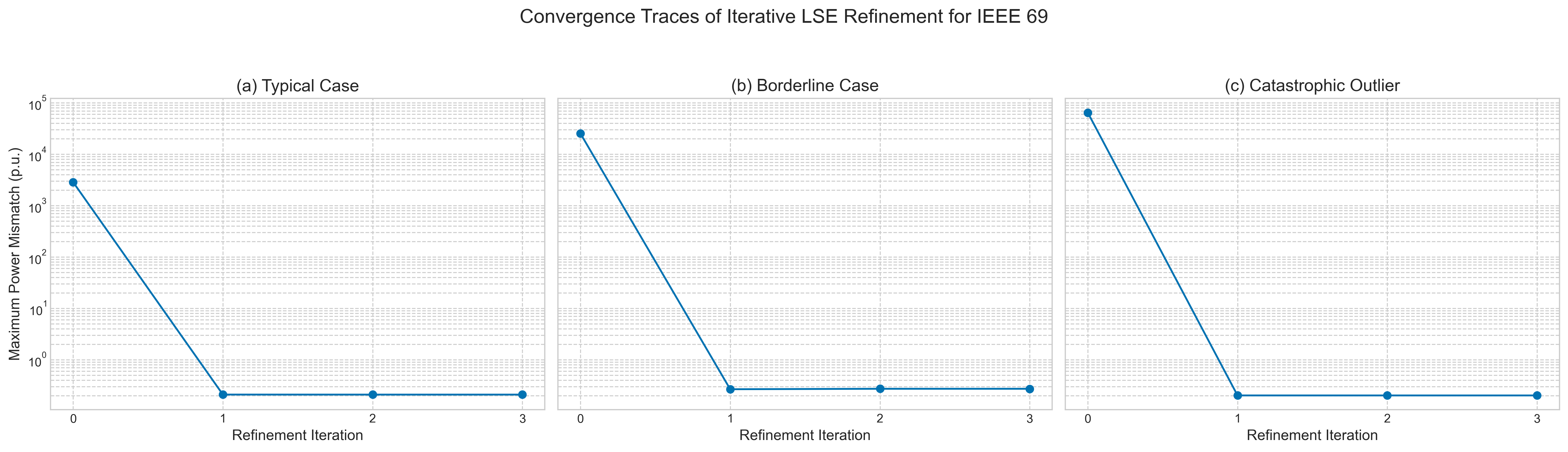}
		\caption{Convergence traces of the iterative LSE refinement for the IEEE 69-bus system. The plots show the maximum power mismatch (log scale) versus the refinement iteration number for (a) a typical well-predicted case, (b) a borderline case, and (c) a catastrophic outlier. Iteration 0 represents the initial raw GNN prediction.}
		\label{fig:convergence_traces}
	\end{figure*}

	To further substantiate the robustness of the LSE refinement, we analyzed its per-iteration convergence behavior across different scenarios. Fig. \ref{fig:convergence_traces} shows the reduction in maximum power mismatch for three representative cases from the IEEE 69-bus test set: a typical case with low initial error, a borderline case, and a catastrophic outlier case. In all scenarios, the LSE refinement demonstrates rapid and stable convergence, reducing the mismatch by orders of magnitude within the first two iterations. This behavior validates the claim that the LSE step is a robust mechanism for enforcing physical consistency, even when starting from a poor initial guess provided by the GNN.
	
	\begin{table}[!htbp]
		{\setlength{\textfloatsep}{8pt plus 2pt minus 2pt}\setlength{\intextsep}{8pt plus 2pt minus 2pt}\setlength{\abovecaptionskip}{6pt}\setlength{\belowcaptionskip}{6pt}
			\setlength{\tabcolsep}{3pt}\renewcommand{\arraystretch}{1.1}}
		\centering
		\caption{Robustness and Generalization Results for the IEEE 118-Bus System.}
		\par\smallskip{All results use a \SI{100}{\mega\volt\ampere} base for IEEE-118 unless otherwise noted.}\label{tab:robustness_results_118}
		\footnotesize
		\sisetup{
			round-mode=places,
			round-precision=2,
			uncertainty-mode=separate,
			table-align-text-after=false,
			exponent-mode = scientific
		}
		\begin{tabularx}{\textwidth}{>{\raggedright\arraybackslash}p{3.0cm} >{\raggedright\arraybackslash}X S[table-format=1.3(3)] S[table-format=1.3(3)]}
			\toprule
			\textbf{Scenario} & \textbf{Metric} & {\textbf{GNN-PINN}} & {\textbf{GNN-PINN+LSE}} \\
			\midrule
			\multirow{6}{*}{\makecell[l]{\textbf{Heavy-Loading}\\\textbf{Stress Test}}}
			& MAE $|V_m|$ (p.u.) & \num{0.116 +- 0.004} & \num{0.067 +- 0.031} \\
			& MAE $V_a$ (deg) & \num{25.06 +- 2.18} & \num{31.96 +- 5.69} \\
			& Max P Mismatch (p.u.) & \num{42.44 +- 3.42} & \num{26.14 +- 6.65} \\
			& Max Q Mismatch (p.u.) & \num{367.24 +- 40.49} & \num{0.62 +- 0.35} \\
			& Avg Time (ms) & {$<0.01$} & \num{3.64 +- 0.72} \\
			& Speedup vs. NR & {$\ge \num[round-precision=2]{1.87e4}$} & \num{3.9} $\times$ \\
			\cmidrule{1-4}
			\multirow{6}{*}{\makecell[l]{\textbf{N-1}\\\textbf{Contingencies}}}
			& MAE $|V_m|$ (p.u.) & \num{0.036 +- 0.008} & \num{0.001 +- 0.002} \\
			& MAE $V_a$ (deg) & \num{5.62 +- 2.54} & \num{5.36 +- 2.55} \\
			& Max P Mismatch (p.u.) & \num{15.98 +- 5.36} & \num{6.13 +- 2.62} \\
			& Max Q Mismatch (p.u.) & \num{22.78 +- 23.32} & \num{0.00 +- 0.05} \\
			& Avg Time (ms) & \num{4.48 +- 0.87} & \num{8.39 +- 1.33} \\
			& Speedup vs. NR & \num{3.2} $\times$ & \num{1.7} $\times$ \\
			\bottomrule
		\end{tabularx}
	\end{table}
	\setlength{\tabcolsep}{6pt}\renewcommand{\arraystretch}{1.0}
	
	\begin{table}[!htbp]
		{\setlength{\textfloatsep}{8pt plus 2pt minus 2pt}\setlength{\intextsep}{8pt plus 2pt minus 2pt}\setlength{\abovecaptionskip}{6pt}\setlength{\belowcaptionskip}{6pt}
			\setlength{\tabcolsep}{3pt}\renewcommand{\arraystretch}{1.1}}
		\centering
		\caption{Robustness and Generalization Results for the IEEE 69-Bus System.}
		\par\smallskip{All results use a \SI{10}{\mega\volt\ampere} base for IEEE-69 unless otherwise noted.}\label{tab:robustness_results_69}
		\footnotesize
		\sisetup{
			round-mode=places,
			round-precision=2,
			uncertainty-mode=separate,
			table-align-text-after=false,
			exponent-mode = scientific
		}
		\begin{tabularx}{\textwidth}{>{\raggedright\arraybackslash}p{3.3cm} >{\raggedright\arraybackslash}X S[table-format=5.2(4)] S[table-format=1.3(3)]}
			\toprule
			\textbf{Scenario} & \textbf{Metric} & {\textbf{GNN-PINN}} & {\textbf{GNN-PINN+LSE}} \\
			\midrule
			\multirow{6}{*}{\makecell[l]{\textbf{Heavy-Loading}\\\textbf{Stress Test}}}
			& MAE $|V_m|$ (p.u.) & \num{0.012 +- 0.001} & {$<0.001$} \\
			& MAE $V_a$ (deg) & \num{23.95 +- 0.01} & {$<0.01$} \\
			& Max P Mismatch (p.u.) & \num{3135.50 +- 706.42} & {$<10^{-8}$} \\
			& Max Q Mismatch (p.u.) & \num{13571.12 +- 1457.07} & {$<10^{-8}$} \\
			& Avg Time (ms) & {$<0.01$} & \num{1.83 +- 0.41} \\
			& Speedup vs. NR & {$\ge \num[round-precision=2]{1.53e4}$} & \num{6.5} $\times$ \\
			\bottomrule
		\end{tabularx}
	\end{table}
	\setlength{\tabcolsep}{6pt}\renewcommand{\arraystretch}{1.0}
	
	\subsection{Implications and Broader Context}
	The significant speedup of our GNN models is operationally significant. Our GNN variants achieve up to \num[exponent-mode = scientific, round-mode=places, round-precision=1]{8.4e3}$\times$ speedup (118-bus, data-only GNN), while our proposed GNN-PINN model achieves a speedup of approximately \num[exponent-mode = scientific, round-mode=places, round-precision=1]{7.0e3}$\times$ on the 118-bus system and \num[exponent-mode = scientific, round-mode=places, round-precision=1]{6.1e3}$\times$ on the 33-bus system. On the 118-bus system, per-case inference time is below our \SI{0.01}{\milli\second} measurement resolution (thus reported speedups are lower bounds). Evaluating 100 contingencies is therefore bounded above by $\le \SI{1}{\milli\second}$ on our hardware. In contrast, a traditional NR solver would require over \SI{1.4}{\second}, failing to meet sub-second operational requirements. Direct numerical comparison with other AI-based methods is challenging due to differences in problem scope (OPF vs. ACPF), test systems, and implementation details. We therefore focus our empirical comparison on established classical solvers (NR, FDPF) that serve as the industry standard. The inclusion of an MLP-PINN baseline (Tables \ref{tab:results_33bus}--\ref{tab:results_118bus}) confirms the advantages of topology-aware models—especially when coupled with LSE—on distribution systems; results are mixed on the 118-bus for raw predictions, while the hybrid GNN-LSE shows the strongest physical consistency at low cost, which consistently achieves lower error and better physical consistency by explicitly modeling the grid topology. Our approach offers a clear advantage over both traditional solvers like NR/FDPF in terms of speed and over pure ML models (like our GNN-DataOnly ablation) in terms of physical reliability. The LSE-style refinement stage acts as a crucial "safety net" that makes data-driven methods more trustworthy for critical applications.
	
	\section{Conclusion}
	This paper introduced a novel hybrid GNN-LSE approach for ACPF, combining a specialized GNN-PINN with a robust, iterative linear refinement step. Comprehensive evaluation on IEEE 33-, 69-, and 118-bus systems demonstrated the model's effectiveness and scalability, achieving up to a \num[exponent-mode = scientific, round-mode=places, round-precision=1]{8.4e3}$\times$ speedup over NR with our GNN variants and a significant reduction in physical power mismatch, particularly after linear refinement. Furthermore, robustness tests confirmed the model's ability to generalize to heavy-loading stress (\SI{120}{\percent}--\SI{150}{\percent} of nominal) and adapt to topological changes from contingencies.
	
	The results highlight a clear and practical trade-off: the GNN-only prediction offers maximum speed for preliminary checks; the linear refinement provides a fast route to a physically-consistent solution suitable for most operational needs; and an optional NR refinement (initialized by the GNN) can provide a physically-exact state with enhanced convergence reliability. A key limitation of this work is its reliance on synthetic data. Future work should focus on bridging the sim-to-real gap using techniques like domain adaptation or training with real-world PMU data. Validating the approach on even larger, more ill-conditioned networks and adapting it for AC-OPF are also promising research directions. Furthermore, the demonstrated speedup makes this framework a prime candidate for application in computationally demanding scenarios such as real-time energy management for microgrids with high penetrations of electric vehicles \cite{41}.
	
	\section*{Declaration of generative AI and AI-assisted technologies in the writing process}
	During the preparation of this work the author(s) used Gemini 2.5 Pro in order to improve the language and clarity of the manuscript. After using this tool/service, the author(s) reviewed and edited the content as needed and take(s) full responsibility for the content of the published article.
	
	\begingroup\sloppy\emergencystretch=3em
	
	\endgroup

	\appendix
	\section{Reproducibility Details}
	\label{sec:hyperparams}
	To ensure the reproducibility of our results, all key hyperparameters used for training the GNN-PINN model and for the linear refinement are detailed in Table \ref{tab:hyperparams}. The initial scaling factors for the loss function components are provided in Table \ref{tab:scaling_factors}.
	
	\begin{table}[!htbp]
		\centering
		\caption{Hyperparameters for GNN Training and Linear Refinement.}
		\label{tab:hyperparams}
		\footnotesize
		\begingroup\renewcommand\tabularxcolumn[1]{m{#1}}
		\begin{tabularx}{\textwidth}{>{\raggedright\arraybackslash}m{4.8cm} >{\raggedleft\arraybackslash}m{3.2cm} >{\raggedright\arraybackslash}X}
			\toprule
			\textbf{Parameter} & \textbf{Value} & \textbf{Description} \\
			\midrule
			\multicolumn{3}{l}{\textbf{GNN Architecture \& Training}} \\
			Optimizer & AdamW & Adaptive learning rate optimizer \\
			Learning Rate & \num{8e-4} & Initial learning rate \\
			LR Scheduler & OneCycleLR & Learning rate scheduler policy \\
			Batch Size & 128 & Number of samples per training batch \\
			Epochs & 500 (max) & Maximum of training epochs \\
			Early Stopping Patience & 40 & Epochs to wait for improvement before stopping \\
			Hidden Dimensions & 128 & Number of channels in hidden GNN layers \\
			Dropout Rate & 0.2 & Regularization dropout rate \\
			Total Learnable Parameters (GNN) & \num{243971} & Total parameters in the GNN model \\
			Total Learnable Parameters (MLP) & \num{373858} & Total parameters in the MLP baseline \\
			Random Seeds & {[}42, 123, 1024, 2048, 4096{]} & Seeds used for the 5 training runs \\
			\midrule
			\multicolumn{3}{l}{\textbf{Loss Function}} \\
			Huber Loss $\delta$ & 1.0 & Threshold for switching between quadratic/linear loss \\
			Dynamic Weight $\beta$ & 0.9 & Momentum parameter for EMA of physics weights \\
			\midrule
			\multicolumn{3}{l}{\textbf{Linear Refinement}} \\
			Max Iterations ($K_{\text{max}}$) & 3 & Number of refinement iterations \\
			\bottomrule
		\end{tabularx}
		\endgroup
	\end{table}

	\begin{table}[!htbp]
		\centering
		\caption[Initial Loss Scaling Factors (s)]{Initial Loss Scaling Factors ($s$) used in Eq. \ref{eq:total_loss}.}
		\label{tab:scaling_factors}
		\footnotesize
		\sisetup{round-mode=places,round-precision=2, exponent-mode = scientific}
		\begin{tabular*}{\textwidth}{@{\extracolsep{\fill}} l S[table-format=1.2] S[table-format=1.2e1] S[table-format=2.2] S[table-format=1.2e-1] S[table-format=1.2e1]}
			\toprule
			\textbf{System} & {\textbf{$s_{\text{data}}$}} & {\textbf{$s_{P}$}} & {\textbf{$s_{Q}$}} & {\textbf{$s_{V}$}} & {\textbf{$s_{S}$}} \\
			\midrule
			IEEE 33-Bus & 7.06 & 1.57 & 0.06 & \num{1.00e-8} & \num{1.20e-1} \\
			IEEE 69-Bus & 10.64 & \num{3.52e6} & 43.63 & \num{1.00e-8} & \num{7.35e7} \\
			IEEE 118-Bus & 7.03 & \num{2.12e5} & \num{9.81e4} & \num{4.20e-1} & \num{1.61e1} \\
			\bottomrule
		\end{tabular*}
		\begin{minipage}{\textwidth}
			\footnotesize\raggedright
			\textit{Note: These values were calculated as the inverse of the initial observed loss magnitudes for each term (Eq.~\ref{eq:total_loss}),}
			\textit{ so that each component initially contributes with comparable scale to the total loss across both the data-driven and physics-informed terms.}
			\vspace{0.25em}
			The significant variance in these factors arises from two primary sources:
			\begin{enumerate}[label*=\arabic*., noitemsep, leftmargin=*]
				\item \textbf{Differences Between Loss Terms (e.g., $s_P, s_Q, s_V, s_S$):} Power residuals can be large at initialization due to the nonlinearity of AC equations, while voltage residuals are typically small.
				\item \textbf{Differences Between Test Cases (e.g., 33/69 vs. 118-bus):} Larger and more meshed systems exhibit higher initial residuals due to denser coupling (e.g., $s_{\text{data}}=7.06$ vs. $10.64$; $s_P=0.06$ vs. $9.81\times 10^4$).
			\end{enumerate}
		\end{minipage}
	\end{table}

	\section{Computational Performance Details}
	\label{sec:comp_perf}
\begin{figure}[!htbp]
		\centering
		\includegraphics[width=0.7\textwidth]{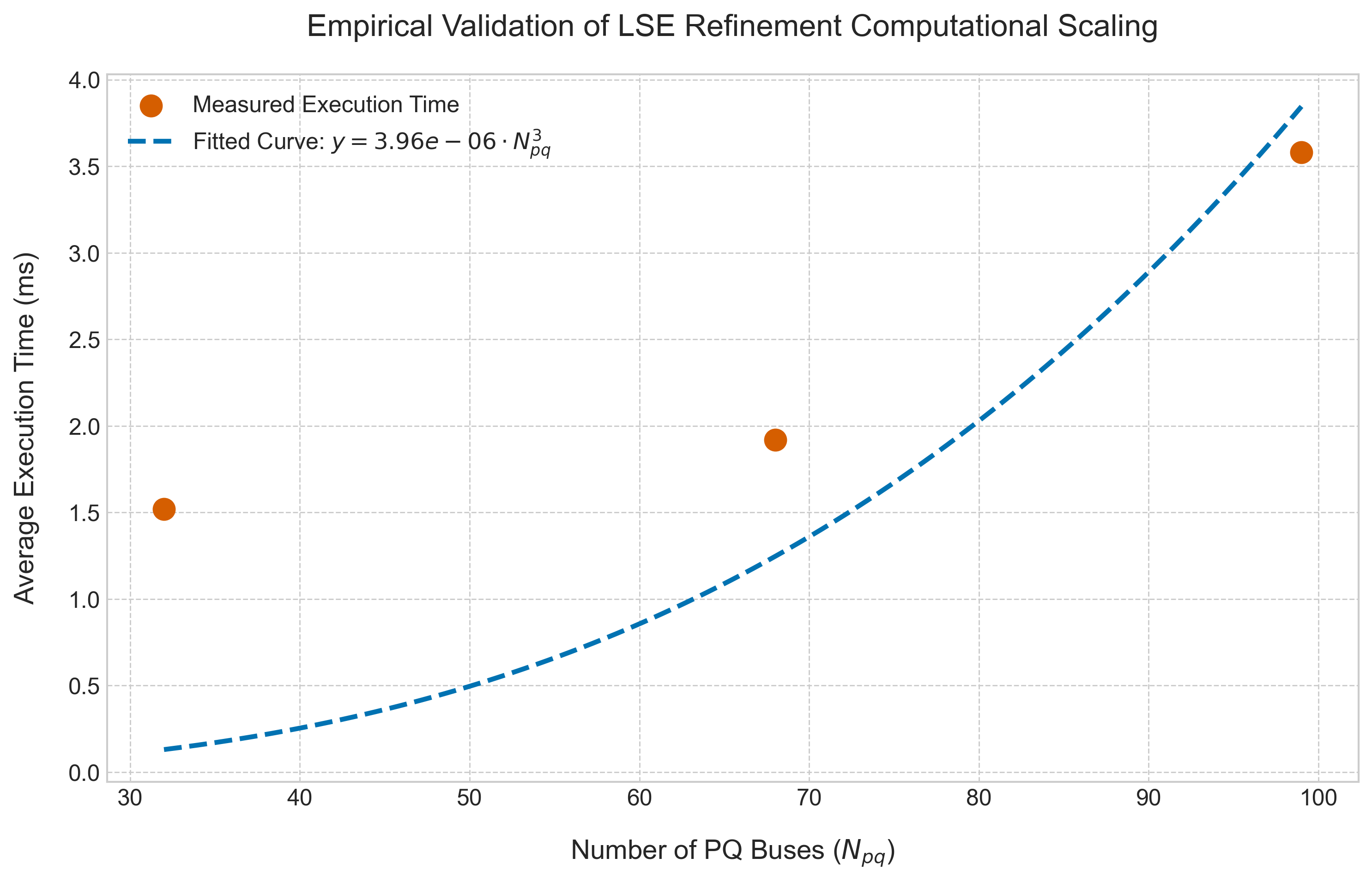}
		\caption[Empirical validation of the LSE refinement complexity]{Empirical validation of the LSE refinement step's computational scaling. The plot shows the measured average execution time vs. the number of PQ buses, with a fitted cubic curve demonstrating the consistency with the theoretical $\mathcal{O}(N_{pq}^3)$ complexity.}
		
		\label{fig:lse_timing_scaling}
	\end{figure}

	\begin{sloppypar}
		\emergencystretch=2em
		To empirically validate the computational complexity of the LSE refinement step, we measured its execution time as a function of the number of PQ buses ($N_{pq}$) for the three test systems. As shown in Fig. \ref{fig:lse_timing_scaling}, the wall-clock time scales consistently with the theoretical $\mathcal{O}(N_{pq}^3)$ complexity associated with solving the dense linear system in Eq. \ref{eq:lse_linear}. While this presents a scaling bottleneck for extremely large systems, the absolute execution time remains very low for the tested networks, confirming its practicality for real-time applications on systems of this scale. To provide further insight into the computational cost, Table \ref{tab:lse_per_iteration_timing} details the per-iteration wall-clock timing statistics, confirming that the computational load is consistent across iterations.
		
	\end{sloppypar}
	
	{%
		\noindent{Because each LSE iteration solves the same dense linear system, the per-iteration wall-clock time is effectively constant; observed differences are at or below the \SI{0.01}{\milli\second} timer resolution.}
		
		\begin{table}[!htbp]
			\centering
			\caption{Per-iteration LSE refinement time and stability across iterations (ms).}
			\label{tab:lse_per_iteration_timing}
			\footnotesize
			\sisetup{round-mode=places,round-precision=2,uncertainty-mode=separate}
			\begin{tabular}{l S[table-format=1.2(2)] S[table-format=1.2]}
				\toprule
				\textbf{System} & {\textbf{Per-iteration (mean $\pm$ std)}} & {\textbf{$\Delta$ across iters}} \\
				\midrule
				IEEE 33-Bus  & \num{0.51 +- 0.11} & 0.01 \\
				IEEE 69-Bus  & \num{0.58 +- 0.15} & 0.00 \\
				IEEE 118-Bus & \num{1.19 +- 0.24} & 0.01 \\
				\bottomrule
			\end{tabular}\\[2pt]
			\footnotesize $\Delta$ is the maximum absolute difference among Iterations 1--3. Times are averaged over 1{,}000 runs with \SI{0.01}{\milli\second} resolution.
		\end{table}
	}
	
\end{document}